%% file: main.tex
\RequirePackage{fix-cm}
\RequirePackage[final]{graphicx}
\documentclass[smallextended, draft]{svjour3}       
\smartqed  

\input{preamble}

\journalname{Empirical Software Engineering}

\begin{document}

\title{The \stackage Repository: An Exploratory Study of its Evolution\footnote{This paper extends a study presented on ACM Symposium on Applied Computing~\cite{legerAl:sac-se2022}.}}




\author{Paul Leger        \and
        Felipe Ruiz \and
        Nicol\'as Sep\'ulveda \and
        Ismael Figueroa \and
        Nicol\'as Cardozo
}


\institute{P. Leger \at
			  Escuela de Ingenier\'ia, 	
              Universidad Cat\'olica del Norte, Chile \\
              \email{pleger@ucn.cl}~Website: \url{http://pragmaticslab.com}           
           \and
           F. Ruiz \at
           Escuela de Ingenier\'ia, \\ 	
              Universidad Cat\'olica del Norte, Chile          
           \and
           N. Sep\'ulveda \at
           Escuela de Ingenier\'ia, 	
              Universidad Cat\'olica del Norte, Chile
           \and
           I. Figueroa \at
           Pragmatics Lab\\ 	
           \url{http://pragmaticslab.com} 
          \and
           N. Cardozo \at
           Systems and Computing Engineering Department, Universidad de los Andes, Colombia\\ 	
           \email{n.cardozo@uniandes.edu.co} 
}

\date{Received: XX/XX/XX   /   Accepted: YY/YY/YY}

\maketitle

\input{abstract}

\input{introduction}

\input{packages}

\input{methodology}

\input{stackage}

\input{answers}

\input{related}

\input{conclusion}




\printbibliography

\end{document}

%% file: preamble.tex
\usepackage[T1]{fontenc}
\usepackage[utf8]{inputenc}

\usepackage{ifdraft}

\usepackage{amsmath}
\usepackage{mathrsfs}
\usepackage{hyperref}
\usepackage[plain]{fancyref}
\usepackage{subcaption}

\usepackage[inline]{enumitem}
\usepackage{xcolor}
\usepackage{tikz}
\usepackage{booktabs}
\usepackage{threeparttable}
\usepackage{multirow}
\usepackage{array}
\usepackage{xspace}
\usepackage[final]{listings}
\usepackage{acronym}
\usepackage{url}
\usepackage{balance}
\usepackage{todonotes}

\usepackage{etoolbox}
\makeatletter
\patchcmd{\@makecaption}
  {\scshape}
  {}
  {}
  {}
\patchcmd{\@makecaption}
  {\\}
  {.\ }
  {}
  {}

\usepackage
  [backend=bibtex,
   natbib=true,
   datamodel=spmpsci,
   hyperref=true,
   url=true,
   defernumbers]{biblatex}

\definecolor{OliveGreen}{rgb}{0,0.6,0.3}

\def\fref{\Fref} 

\renewcommand*{\fancyrefenumlabelprefix}{enum}
\renewcommand*{\Frefenumname}{}
\renewcommand*{\frefenumname}{\MakeLowercase{\Frefenumname}}
\Frefformat{vario}{\fancyrefenumlabelprefix}%
  {\Frefenumname\fancyrefdefaultspacing#1#3}
\frefformat{vario}{\fancyrefenumlabelprefix}%
  {\frefenumname\fancyrefdefaultspacing#1#3}
\Frefformat{plain}{\fancyrefenumlabelprefix}%
  {\Frefenumname\textbf{#1}}
\frefformat{plain}{\fancyrefenumlabelprefix}%
  {\frefenumname\textbf{#1}}    

\renewcommand{\lstlistingname}{Snippet}
\newcommand*{\fancyreflstlabelprefix}{lst}
\newcommand*{\Freflstname}{\lstlistingname}
\newcommand*{\freflstname}{\MakeLowercase{\lstlistingname}}
\Frefformat{vario}{\fancyreflstlabelprefix}%
  {\Freflstname\fancyrefdefaultspacing#1#3}
\frefformat{vario}{\fancyreflstlabelprefix}%
  {\freflstname\fancyrefdefaultspacing#1#3}
\Frefformat{plain}{\fancyreflstlabelprefix}%
  {\Freflstname\fancyrefdefaultspacing#1}
\frefformat{plain}{\fancyreflstlabelprefix}%
  {\freflstname\fancyrefdefaultspacing#1}

\newcommand*{\fancyreflnlabelprefix}{ln}
\newcommand*{\Freflnname}{Line}
\newcommand*{\freflnname}{\MakeLowercase{\Freflnname}}
\Frefformat{vario}{\fancyreflnlabelprefix}%
  {\Freflnname\fancyrefdefaultspacing#1#3}
\frefformat{vario}{\fancyreflnlabelprefix}%
  {\freflnname\fancyrefdefaultspacing#1#3}
\Frefformat{plain}{\fancyreflnlabelprefix}%
  {\Freflnname\fancyrefdefaultspacing#1}
\frefformat{plain}{\fancyreflnlabelprefix}%
  {\freflnname\fancyrefdefaultspacing#1}    

\newcommand*{\fancyrefdeflabelprefix}{def}
\newcommand*{\Frefdefname}{Definition}
\newcommand*{\frefdefname}{\MakeLowercase{\Frefdefname}}
\Frefformat{vario}{\fancyrefdeflabelprefix}%
  {\Frefdefname\fancyrefdefaultspacing#1#3}
\frefformat{vario}{\fancyrefdeflabelprefix}%
  {\frefdefname\fancyrefdefaultspacing#1#3}
\Frefformat{plain}{\fancyrefdeflabelprefix}%
  {\Frefdefname\fancyrefdefaultspacing#1}
\frefformat{plain}{\fancyrefdeflabelprefix}%
  {\frefdefname\fancyrefdefaultspacing#1}    
  
\newcommand*{\fancyreftheolabelprefix}{theo}
\newcommand*{\Freftheoname}{Theorem}
\newcommand*{\freftheoname}{\MakeLowercase{\Freftheoname}}
\Frefformat{vario}{\fancyreftheolabelprefix}%
  {\Freftheoname\fancyrefdefaultspacing#1#3}
\frefformat{vario}{\fancyreftheolabelprefix}%
  {\freftheoname\fancyrefdefaultspacing#1#3}
\Frefformat{plain}{\fancyreftheolabelprefix}%
  {\Freftheoname\fancyrefdefaultspacing#1}
\frefformat{plain}{\fancyreftheolabelprefix}%
  {\freftheoname\fancyrefdefaultspacing#1}    

\newcommand*{\fancyrefproplabelprefix}{prop}
\newcommand*{\Frefpropname}{Proposition}
\newcommand*{\frefpropname}{\MakeLowercase{\Frefpropname}}
\Frefformat{vario}{\fancyrefproplabelprefix}%
  {\Frefpropname\fancyrefdefaultspacing#1#3}
\frefformat{vario}{\fancyrefproplabelprefix}%
  {\frefpropname\fancyrefdefaultspacing#1#3}
\Frefformat{plain}{\fancyrefproplabelprefix}%
  {\Frefpropname\fancyrefdefaultspacing#1}
\frefformat{plain}{\fancyrefproplabelprefix}%
  {\frefpropname\fancyrefdefaultspacing#1}

\lstdefinestyle{floating}
 {
  float=htbp,
  captionpos=b,
  numberstyle=\scriptsize\color{gray},
  numbersep=-2pt,
  stepnumber=1,
  belowskip=-1\baselineskip,
  aboveskip=-0.3\baselineskip,
  escapechar=`
}

\lstdefinelanguage{JavaScript}{
keywords={},
keywordstyle=\color{purple}\bfseries,
ndkeywords={typeof, new, true, false, catch, function, return, null, catch, switch, var, if, in, for, while, do, else, case, break, throw, this, instanceof},
ndkeywordstyle=\bfseries,
identifierstyle=\color{black},
sensitive=false,
comment=[l]{//},
morecomment=[s]{/*}{*/},
stringstyle=\color{green!50!brown}\ttfamily,
morestring=[b]',
morestring=[b]"
}

\lstdefinestyle{ctxtraits}
 {language=JavaScript,
  frame=lines,
  showstringspaces=false,
  tabsize=3,
  style=floating,
  morekeywords={Trait, cop, Context, activate, deactivate, adapt, addObjectPolicy, manager}
}

\lstnewenvironment{ctxtraits}[1][]
 {\lstset{style=ctxtraits,#1}}{}  
 
\tikzstyle{place}=[circle,thick,draw=black!75,minimum size=5mm]
\tikzstyle{iplace}=[circle,dashed,thick,draw=black!75,minimum size=5mm]
\tikzstyle{itransition}=[rectangle,draw,thick,fill=black,minimum size=1mm]
\tikzstyle{etransition}=[rectangle,draw,thick,minimum size=1mm]
\tikzstyle{ctransition}=[rectangle,draw,color=black!45,thick,fill=black!45,minimum size=1mm]

\newcommand{\eg}{\emph{e.g.,}\xspace}
\newcommand{\ie}{\emph{i.e.,}\xspace}

\newcommand{\aka}{\emph{a.k.a.}\xspace}
\newcommand{\parhead}[1]{\noindent{\bf {\em #1.}}}

\newcommand{\mtl}[0]{\mbox{\co{mtl}}\xspace}
\newcommand{\transformers}[0]{\mbox{\co{transformers}}\xspace}
\newcommand{\free}[0]{\mbox{\co{free}}\xspace}
\newcommand{\monadcontrol}[0]{\mbox{\co{monad-control}}\xspace}

\newcommand{\haskell}[0]{\mbox{Haskell}\xspace}
\newcommand{\hackage}[0]{\mbox{Hackage}\xspace}
\newcommand{\stackage}[0]{\mbox{Stackage}\xspace}
\newcommand{\javascript}[0]{\mbox{JavaScript}\xspace}

\newcommand{\python}[0]{\mbox{Python}\xspace}
\newcommand{\rust}[0]{\mbox{Rust}\xspace}
\newcommand{\cran}[0]{\mbox{CRAN}\xspace}
\newcommand{\ruby}[0]{\mbox{Ruby}\xspace}
\newcommand{\npm}[0]{\mbox{NPM}\xspace}

\newcommand{\co}[1]{\mbox{{\fontfamily{phv}\selectfont\small\sffamily  #1}\xspace}}

\definecolor{author}{rgb}{.5, .5, .5}
\definecolor{comment}{rgb}{.1, .0, .9}
\definecolor{note}{rgb}{.9, .4, .0}
\definecolor{idea}{rgb}{.1, .7, .0}
\definecolor{missing}{rgb}{.9, .1, .0}

\input{acronyms}

\let\orig@figure\figure
\renewcommand*{\figure}[1][]{\orig@figure[#1]\vspace{-1ex}} 
\let\orig@endfigure\endfigure
\renewcommand*{\endfigure}{\vspace{-1ex}\orig@endfigure} 

\usepackage{caption}
\makeatother

%% file: acronyms.tex
\acrodef{MSR}{Mining Software Repositories}
\acrodef{AOP}{Aspect-oriented Programming}
\acrodef{API}{Application Programming Interface}
\acrodef{AST}{Abstract Syntax Tree}
\acrodef{COP}{Context-oriented programming}
\acrodef{DSL}{Domain Specific Language}
\acrodef{IOT}[IoT]{Internet of Things}
\acrodef{IR}{Intermediate Representation}
\acrodef{SMT}{Satisfiability Modulo Theory}

\newcommand{\acResetNonTrivial}
  {\acresetall
   \acused{CPU}
   \acused{AST}
   \acused{API}
   \acused{LAN}
   \acused{SMT}
   \acused{GUI}}

\acResetNonTrivial

%% file: abstract.tex

\begin{abstract}~\\
\parhead{Context} Package repositories for a programming language are increasingly common. A repository can keep a register of the evolution of its packages. In the programming language \haskell, with its defining characteristic monads, we can find the \stackage repository, which is a curated repository for stable \haskell packages in the \hackage repository. Despite the widespread use of \stackage in its industrial target, we are not aware of much empirical research about how this repository has evolved, including the use of monads.

\parhead{Objective} This paper conducts empirical research about the evolution of \stackage considering monad packages through 22 Long-Term Support releases during the period 2014-2023. Focusing on five research questions, this evolution is analyzed in terms of packages with their dependencies and imports; including the most used monad packages. To the best of our knowledge, this is the first large-scale analysis of the evolution of the \stackage repository regarding packages used and monads.  
      
\parhead{Method} We define six research questions regarding the repository's evolution, and analyze them on 51,716 packages (17.05 GB) spread over 22 releases. For each package, we parse its cabal file and source code to extract the data, which is analyzed in terms of dependencies and imports using Pandas scripts.
 
\parhead{Results} From the methodology we get different findings. For example, there are packages that depend on other packages whose versions are not available in a particular release of \stackage; opening a potential stability issue. The \mtl and \transformers are on the top 10 packages most used/imported  across releases of the \stackage evolution. We discussed these findings with \stackage maintainers, which allowed us to refine the research questions.    

\parhead{Conclusions} On the one hand, like previous studies, these results may evidence how developers use \haskell and give guidelines to \stackage maintainers. One of our proposals is to generate control over the categories and stability that developers assign to their packages. On the other hand, we recommend that \stackage designers take more care when verifying the versions of package dependencies.

\keywords{\stackage \and \haskell \and Mining Software Repository}
\end{abstract}

%% file: introduction.tex

\section{Introduction}
\label{sec:intro}

\haskell~\cite{Haskell-lang} is a well known {\em pure} functional programming 
language due to its explicit representation of  {\em side-effects} within the language's type 
system (\eg input/output expressions, mutable state, exceptions).  To allow developers to have the 
benefits of side-effects, \haskell provides the {\em monad} 
abstraction~\cite{moggi:ic1991,wadler:lfp1990,wadler:afp1995}, a denotational approach to embed 
and reason about {\em notions of computation} such as mutable state, or exceptions. Monads offer 
a strong theoretical foundation to represent side-effects. The \haskell community uses two widely 
known package archive repositories of open source software: \hackage~\cite{hackage} and 
\stackage~\cite{stackage}. The former is the central package for \haskell, the latter is a curated 
repository of \hackage packages. To date, August 2023, the \stackage community is managing 22 
Long-Term Support~(LTS) releases, dating from 2014. Additionally, \haskell developers
have downloaded over 85 million packages from the \stackage selection. 

To the best of our knowledge, although \stackage is the main repository selected for \haskell 
packages, the main focus of study on \haskell repositories has been around the use of abstractions in 
\hackage~\cite{figueroa2017,figueroaAl:scp2020,morris:haskell2010}. Currently, there are no studies  
focusing on how the package repository evolves, neither on the use of monads. As this is still an open 
question, this paper focuses on several quantitative results obtained from an empirical investigation 
about the evolution of the \stackage repository. 

As \stackage is a package selection of \hackage, this repository uses the {\em Cabal} system~\cite{Cabal}, which we can use as a tool to directly download, build, and install packages from \hackage. Using the Cabal system and parsing source code, we can collect information as descriptions and package dependencies to carry out this study. This information allows us to perform an exploratory and quantitative study that consists of a massive analysis of packages present in 22 releases (17.05 gigabytes). In addition to describing \stackage's evolution and monad packages, we answer the following Research Questions (RQs) per release:   

\begin{enumerate}[label=\textbf{RQ\arabic*}, leftmargin=1.3cm]
  \item \label{enum:imported}
  Which packages are imported the most by other \stackage packages? Do these packages have 
  {\em unstable} or {\em incompatible} dependencies according to their \stackage release? 
  \item \label{enum:dependencies}
  What is the average number of (in)direct dependencies per package?
  \item \label{enum:frequency}
  How frequently are packages updated? 
\end{enumerate}

Additionally, we selected the top most imported packages that provide monads in \stackage, which 
are: \mtl, \transformers, \monadcontrol, and \free. We briefly describe these packages and how they 
are imported. Additionally, we analyze the evolution of dependencies of monad modules available for 
them to answer the following research questions: 

\begin{enumerate}[label=\textbf{RQ\arabic*}, leftmargin=1.3cm]
\setcounter{enumi}{3}
  \item \label{enum:monads-evol}
  How have the selected monad packages evolved? 
  \item \label{enum:monads-use}
  How has the use of the selected monad packages evolved?
  \item \label{enum:mtl-transformers-dependencies}
  How many packages that depend on the \mtl and \transformers packages are added to 
  and removed from \stackage? How many packages that depended on these monad packages stopped 
  their dependencies?  
\end{enumerate}

By answering these research questions, this paper provides language researchers and \stackage 
administrators with empirical information for the design and development of policies for language 
repositories. From this research result, we can provide some insightful findings as examples. First, a 
growing trend of packages:
\begin{enumerate*}[label=(\arabic*)] 
\item depend on other packages whose versions are not available in its particular release, 
\item have dependencies with incompatible versions, and
\item need that the developers implement future revisions of the package dependencies to fix (not always) the previous two issues.    
\end{enumerate*}
This finding is re-analyzed with information provided by the \stackage maintainers through a public 
GitHub forum. Second, the \mtl and \transformers packages are among the top ten most used packages in \stackage. 
Third, although \mtl and \transformers are widely used, there is {\em not} a growing trend to their use. This paper extends a study presented on ACM Symposium on Applied Computing (SAC'2022)~\cite{legerAl:sac-se2022} in the following aspects:

\begin{enumerate}
	\item Increase the number of \stackage releases analyzed (57.1\% more): 22 releases instead of 14.
	\item Add and refine research questions. In specific, RQ4 is new, and RQ5 with RQ6 are refined through analyzing new packages that implement monads.    
	\item Add new monads packages analyzed. Apart from the \mtl package, we add \transformers, \monadcontrol, and \free. We analyzed the import of monad packages to select these new packages.
	\item Add a discussion of the answers to these research questions with \stackage maintainers to refine some answers.
	\item An automatized set of \python scripts to download and process the \stackage releases. Using Jupiter\footnote{\url{http://jupyter.org}}, a web-based interactive computing platform for different programming languages, these research questions (and potentially others) can be answered. This script is available on \url{https://github.com/pragmaticslaboratory/stackage-evolution} (rev. \co{6065acc})~\cite{scriptStackage}.       
\end{enumerate}
     
The rest of this paper is organized as follows. \fref{sec:bk} briefly presents concepts of the evolution of package repositories and monads. Section~\ref{sec:methodology} introduces the methodology used for this study. Section~\ref{sec:sk} describes the \stackage evolution through different graphs. We then answer and discuss research questions in Section~\ref{sec:results}. Additionally, this section revised the second research question using information provided by the \stackage maintainers. Section~\ref{sec:rw} discusses the related work, and Section~\ref{sec:conc} concludes with future work. 

\smallskip


%% file: packages.tex

\section{Evolution of Package Repository and Monads}
\label{sec:bk}

This section presents the core concepts needed to understand this study. Firstly, we introduce concepts of package repositories, Mining Software Repositories (MSR)~\cite{hemmatiAll:msr2013}, and repository evolution. Secondly, we introduce monads, the main packages offering monad functionality, and their uses.  

\subsection{Package Repository Evolution}
\label{sec:pre}

Package repositories allow developers to store, access, and share packages (\eg 
GitHub\footnote{\url{http://github.com}}). Some repositories are specialized for one programming 
language. Using specialized repositories, a developer community of a language can create, access, 
and share packages. As examples, we can find NPM~\cite{npm} for \javascript, PYP~\cite{pyp} for 
\python, NuGet~\cite{nuget} for .Net, and \hackage~\cite{hackage} for \haskell.             


A repository can evolve with different releases over time, where each release contains a list of 
selected packages with specific versions and does not potentially have conflict dependencies among 
them. In \haskell, developers can use \stackage~\cite{stackage}, a repository of curated \haskell 
packages that are selected from \hackage. Additionally, \stackage keeps a set of Long-Term Support~(LTS) 
releases since 2014. Apart from repositories for programming languages, we can find package 
repositories in other domains, like Linux distributions 
(\eg Debian\footnote{\url{http://wiki.debian.org/DebianReleases}}) or software development 
(\eg GitLab\footnote{\url{http://docs.gitlab.com/ee/user/project/releases}}). 

In MSR -- a field that analyzes the data available in software repositories -- empirical studies have allowed 
developers and researchers to gather insight and build a general panorama of a 
repository~\cite{Decan2018,Decan2019,Decan2015,huAl:springplus2016,kikasAl:msr2017,munaiahAl:ese2017,Plakidas2017}. For these studies, different metrics and tools are used to analyze in detail how 
repositories evolve or are distributed. For example, ~\citet{Plakidas2017} conduct a study on the 
evolution of a repository for the R programming language using the package metadata. Similarly, there 
is an analysis identifying the \emph{technical debt} among NPM package dependencies (\ie outdated 
dependencies)~\cite{Decan2018, gonzalezAl:oss2017}. ~\citet{kikasAl:msr2017} analyze the 
dependency network of software repositories for \javascript, Ruby, and Rust. Finally, 
~\citet{huAl:springplus2016} explore the {\em influence} of GitHub repositories from their users 
(developers) between 2012 and 2015, where the influence is measured by the number of {\em forks} 
and {\em stars} that a repository contains.      

\subsection{Monads in \haskell}
\label{sec:mh}

Monads are defined as abstractions that allow developers to encapsulate and reason about 
{\em notions of computations}, such as mutable state, I/O, or exceptions through a denotational 
approach. Monads were originally introduced in \haskell~\cite{moggi:1988,wadler:lfp1990}, and 
later extended to the concept of monad \transformers~\cite{liang:popl95}. Additionally, monads put 
forward theoretical foundations~\cite{moggi:ic1991} that have even been applied in other domains 
such as Aspect-Oriented Programing~\cite{demeuter:ecoop1997}, and more recently Context-Oriented 
Programming~\cite{aotaniAl:COP2015}. Furthermore, to simplify the use of monads, they have been 
introduced as a {\em design pattern}~\cite{GoF94}. 

In the \haskell community, we can find a number of packages that provide monads with standard or 
custom implementation such as: 
\transformers, the {\em monad transformer library} (\aka \mtl), 
\monadcontrol, \free, \co{monads-tf}, or \co{monads-fd}.  In this study, we focus on the former four 
monad packages, which constitute the main packages used in the \haskell community.

The \transformers package implements the functor and monad functionality as a portable library used 
to lift state transformers into higher-order functions that can be used generally by 
programs~\cite{jones:afp1995}.
The \mtl package builds on the ideas of \transformers, providing a standardized and flexible use of 
monads~\cite{liang:popl95} through the introduction of the \co{Monad} class. Using monad classes, it 
is possible to have more generic lifting of the monadic behavior through functional dependencies.
The \monadcontrol package is used to lift generic control operations existing in other monads 
(\eg \co{IO}). 
Finally, the \free package offers unrestricted monads to use in managing tree-like structures or the 
implementation of Domain Specific Languages (DSLs).

There are different studies regarding the use of monads in \haskell. ~\citet{Algehed2017} show how 
monads can be used to statically track data dependencies. ~\citet{bentonAl:appsem2002} explain 
the impact of monads in terms of applications in pure functional languages like \haskell. Finally, 
~\citet{Jaskelioff2011} present the \co{monatron} monad to address flexibility and expressiveness 
issues in existing monads. \fref{sec:rw} presents a deeper discussion about this topic.

%% file: methodology.tex

\section{Methodology}
\label{sec:methodology}

\begin{figure*}[ht]
  \centering
  \includegraphics[width=1\linewidth]{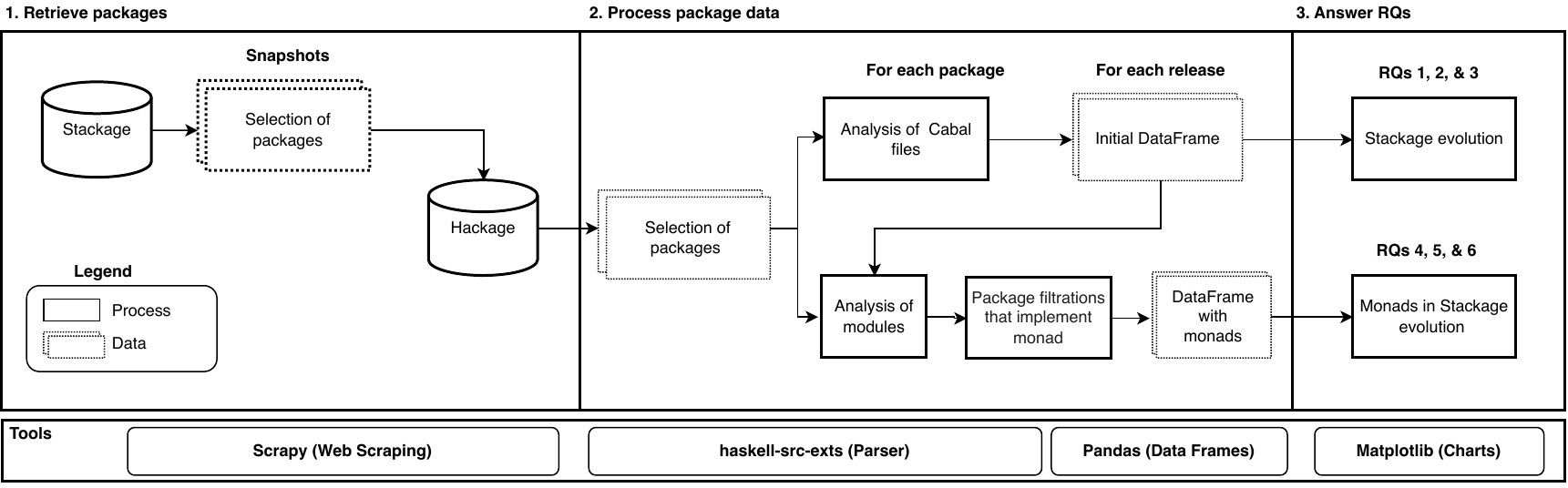}
  \caption{Workflow of our methodology}
  \label{fig:methodology}
\end{figure*}

Following guidelines of MSR methodologies~\cite{hemmatiAll:msr2013} and similar 
studies~\cite{figueroa2017,figueroaAl:scp2020}, we define the methodology for our study. 
\fref{fig:methodology}, shows the three stages composing our methodology: 
\begin{enumerate*}[label=(\arabic*)]
\item retrieving packages, 
\item processing package data, and 
\item answering to the research questions. 
\end{enumerate*} 

The execution of each stage is supported by its own set of tools. The 
{\em Scrapy}~\cite{scrapy} tool is used to download packages with available download links 
on \stackage. To analyze packages, we use {\em Pandas}~\cite{pandas}, and 
{\em Matplotlib}~\cite{matplotlib}. Finally, we use the \emph{haskell-src-exts} (HSE) package to parse 
package modules. In the following we describe each stage of our methodology.

\subsection{Retrieving Packages}
\label{sec:mpo}

Our study considers all 22 releases published by \stackage between 2014 and 2023, where each release 
contains around 2,000 packages. Using Scrapy, we automatically download all \haskell packages from all releases. Scrapy provides 
\emph{Spiders} that allow developers to define how to extract information from a Website. We start 
retrieving the package list of a specific \stackage release. As \stackage is a package selection of 
\hackage, Scrapy directly downloads packages from \hackage. The average time of retrieving the packages for each realease, following this process, is one hour.

\subsection{Processing Package Data}
\label{sec:mppd}

To process package data, we use \co{DataFrames} in Pandas, given its flexibility to (re)structure data. 
In each package, there is a \co{.Cabal} file that contains its manifest. We use a \haskell implemented script to statically analyze these files and obtain meta-data such as: package dependencies, categories, and the provided modules. To clarify the meta-data processed, we briefly describe the fields that allow us to answer the research questions:
\begin{description}
\item[{\bf Release:}] A unique string in \stackage, which indicates the release version.
\item[{\bf Name:}] A unique string in \hackage. We can find the same package in
different releases.
\item[{\bf Version:}] A string that follows the format \co{X.Y.Z} (Major.Minor.Patch). We
can find the same package with different versions in the releases.
\item[{\bf Stability:}] Free string without a format, which can be added by the developer. This field reports the stability of the package.
\item[{\bf Category:}] A list of free strings that describe the categories to which the
package belongs. This list must be defined by the developer.
\item[{\bf Dependencies:}] List of package names and their version ranges.
\item[{\bf Provided modules:}] List of modules publicly exposed by the package.
\item[{\bf Main modules:}] List of executable modules in the package. A package
can specify multiple executables each with a main module.
\end{description} 

With the retrieved meta-data, we analyzed the packages that implement monadic functionality. In 
particular, we inspect the dependencies of each package to quantify how and which monad 
packages are required. To achieve this, we use a \haskell script with HSE 
to allow us to parse the actions used from each of the selected monad packages: \mtl, \transformers, \monadcontrol, and \free. To select these monad packages, we followed these three steps:
 
 \begin{enumerate}[label=\textbf{Step \arabic*.}, leftmargin=1.3cm]
 \item We executed a script that revised the title and descriptions of each package that contains the word ``monad'' and similar variants. This script returned a list of potential monad packages.
 \item We manually filtered the list from Step 1., to obtain the packages that actually implement monadic behavior.
 \item With the filtered list, we run a script that returns the packages that had been imported more than 0.1\% from the total number of imports in a release.       	 	
 \end{enumerate}

Considering these monad packages, we add new fields to the \co{DataFrames} that describe the monad modules used by each package. With this, we can label and count the use of each monad from the different monad libraries. Each added field corresponds to the use of a specific monad module from the library. We add two fields general for all four packages:

\begin{description}
  \item[{\bf Imported modules:}] A set of package names that are imported into the source code of each 
    package. Each module appears only once, even if it is imported into other files.
  \item[{\bf Module-direct flag:}] This flag indicates whether the package depends on one of the 
    four monad packages (\ie \transformers, \mtl, \monadcontrol, or \free). That is, if one of the four monad 
    packages appears in its dependencies field.
\end{description}

\begin{table}[t]
  \centering
  \caption{Fields added to gather data about the use of required monad packages}
  \label{tab:flags}
  \input{joint-flags}
\end{table}

For each package, we add new specific fields that provide information about the monads used from 
each of them. Each field is associated with the monad provided by the packages, as shown in 
\fref{tab:flags}. The uses of these monads are counted whenever a monad is imported into a package.

\subsection{Answering the Research Questions}
\label{sec:marq}

We now turn our attention to the information sources required to answer the six research questions 
posit to understand the evolution and use of \stackage and monad packages. To answer these 
questions, all repository data was structured and processed in \co{DataFrames} using Pandas. We 
briefly describe the process carried out for each question using these data frames. 

\subsubsection{\fref{enum:imported}. Which packages are imported the most by other \stackage 
packages? Do these packages have unstable or incompatible dependencies according to their 
\stackage release?}
\label{sec:marq1}

Using the {\em dependencies} field, we created a list of dependencies (with their versions) for each package in a release. According to the dependencies, we created a dependency network graph, where it is possible to observe the top 10 packages most used/imported in some releases of the \stackage evolution. Additionally, we found out the (number of) packages that depend on versions of packages that are not available in its specific \stackage release. Finally, using the version information per package, we discovered incompatible dependency versions per release, \ie at least two packages that depend on the same package with no intersections of ranges in versions. 

\subsubsection{\fref{enum:dependencies}. What is the average number of (in)direct dependencies 
per package?} 
\label{sec:marq2}

We carried out a {\em Depth First Search (DFS)}~\cite{tarjan1972depth} in package dependencies 
for each release used in this study. This DFS allowed us to quantify the average of (in)direct 
dependencies per package. Additionally, this analysis also allowed us to determine how representative 
these averages are. This analysis of the package dependency is relevant to show that 
\begin{enumerate*}[label=(\arabic*)]
\item the repository dependency, and 
\item the risk that a package can have conflicts when some of its dependencies are removed from a 
specific release or discounted by authors. 
\end{enumerate*}
Likewise, there is a high average pf positive effects in dependencies, for example, package 
reuse and community support to improve maintainability.

\subsubsection{\fref{enum:frequency}. How frequently are packages updated?} 
\label{sec:marq3}

Addressing this question involves counting the number of packages that updated their versions for 
each release. Then, we compare the number of {\em updated} packages to the number of all 
packages in a release. Using this result, we can have a proxy for {\em technical lag}~\cite{gonzalezAl:oss2017} in dependencies, which is helpful to know how frequently bug fixes and 
new features are included in packages.

\subsubsection{\fref{enum:monads-evol}. How have the selected monad packages evolved?} 
\label{sec:marq4}

For this question, we parsed the source code of each package, across all releases, to count 
which packages import at least one of the four monad packages: \mtl, \transformers, \monadcontrol, or 
\free. With this outcome, we can have a sense of the weight of monad behavior in the functionality of 
\stackage packages.

\subsubsection{\fref{enum:monads-use}. How has the use of the selected monad packages evolved?} 
\label{sec:marq5}

To answer this question, we count the modules that are imported from the four selected monad 
packages through all releases. Then, we compare the number of packages that import these modules 
to the number of all packages for one selected monad package. With this information, we can get a 
finer analysis of how monad packages are used, and included as modules for other packages.

\subsubsection{\fref{enum:mtl-transformers-dependencies}. How many packages that depend on the \mtl and \transformers packages are added to and removed from \stackage? How many packages that depended on these monad packages stopped their dependencies?}
\label{sec:marq6}

To answer the first part of this question, we count each package in a release, added and removed, 
that imported \mtl and \transformers modules in a release. For the second part, we followed a similar 
process, in this case counting packages that used \mtl and \transformers modules and then stopped 
using them. In this question, we did not analyze the \monadcontrol and \free because its participation 
in each release is not so high.

%% file: joint-flags.tex
\resizebox{1.01\columnwidth}{!}{
\begin{tabular}{c | l}
\textbf{Field} & \textbf{Monads used} \\
\toprule
\multicolumn{2}{l}{\textbf{\mtl package}}\\
\midrule
Cont & \co{Control.Monad.Cont}, \co{Control.monad.Cont.Class} \\

Error &  \co{Control.Monad.Error}, \co{Control.Monad.Error.Class} \\

Identity & \co{Control.Monad.Identity} \\

List & \co{Control.Monad.List} \\

RWS & \co{Control.Monad.RWS}, \co{Control.Monad.RWS.Class}, \co{Control.Monad.RWS.Lazy}, \co{Control.Monad.RWS.Strict} \\

Reader & \co{Control.Monad.Reader}, \co{Control.Monad.Reader.Class} \\

Writer & \co{Control.Monad.Writer}, \co{Control.Monad.Writer.Class}, \co{Control.Monad.Writer.Lazy}, \co{Control.Monad.Writer.Strict} \\

State & \co{Control.Monad.State}, \co{Control.Monad.State.Class}, \co{Control.Monad.State.Lazy},  \co{Control.Monad.State.Strict} \\

Trans & \co{Control.Monad.Trans}, \co{Control.Monad.Trans.Class} \\

Except & \co{Control.Monad.Except} (available only since version 2.2.1) \\
\midrule
\multicolumn{2}{l}{\textbf{\transformers package}} \\
\midrule
Maybe &  \co{Control.Trans.Maybe} \\

Cont &  \co{Control.Trans.Cont}  \\

Class & \co{Control.Trans.Class} \\ 

RWS & \co{Control.Monad.Trans.RWS}, \co{Control.Monad.Trans.RWS.CSP}, \co{Control.Monad.Trans.RWS.Lazy}, \co{Control.Monad.Trans.RWS.Strict} \\

Identity & \co{Control.Trans.Identity} \\

Except & \co{Control.Trans.Except}  \\

Writer & \co{Control.Monad.Trans.Writer}, \co{Control.Monad.Trans.Writer.CPS},  \co{Control.Monad.Trans.Writer.Lazy},  \co{Control.Monad.Trans.Writer.Strict} \\

Reader & \co{Control.Monad.Trans.Reader} \\

State & \co{Control.Monad.Trans.State}, \co{Control.Monad.Trans.State.Lazy}, \co{Control.Monad.Trans.State.Strict} \\
\midrule
\multicolumn{2}{l}{\textbf{\monadcontrol package}} \\
\midrule
Monad-control & \co{Control.Monad.Trans.Control} \\
\midrule
\multicolumn{2}{l}{\textbf{\free package}} \\
\midrule
Alternative & \co{Control.Alternative.Free}, \co{Control.Alternative.Free.Final} \\

Applicative &  \co{Control.Applicative.Free}, \co{Control.Applicative.Free.Fast}, \co{Control.Applicative.Free.Final} \\

A.Trans & \co{Control.Applicative.Trans.Free} \\

Comonad & \co{Control.Comonad.Cofree}, \co{Control.Comonad.Cofree.Class}, \co{Control.Comonad.Trans.Cofree}, \co{Control.Comonad.Coiter} \\

Free & \co{Control.Monad.Free}, \co{Control.Monad.Free.Ap}, \co{Control.Monad.Free.Church}, \co{Control.Monad.Free.Class}, \co{Control.Monad.Free.TH} \\

M.Trans & \co{Control.Monad.Trans.Free}, \co{Control.Monad.Trans.Free.Ap}, \co{Control.Monad.Trans.Free.Church}, \co{Control.Monad.Trans.Iter} \\
\bottomrule
\end{tabular}
}

%% file: stackage.tex

\section{Stackage: A Brief Overview}
\label{sec:sk}

\stackage~\cite{stackage} is a repository of curated \haskell packages that come from the central 
package repository, \hackage~\cite{hackage}. Between 2014 and 2023, the \stackage community 
released 22 Long-Term Support~(LTS) releases. Before answering our research questions, we briefly 
describe this repository.   

\begin{figure*}[ht]
\centering\includegraphics[width=1\linewidth]{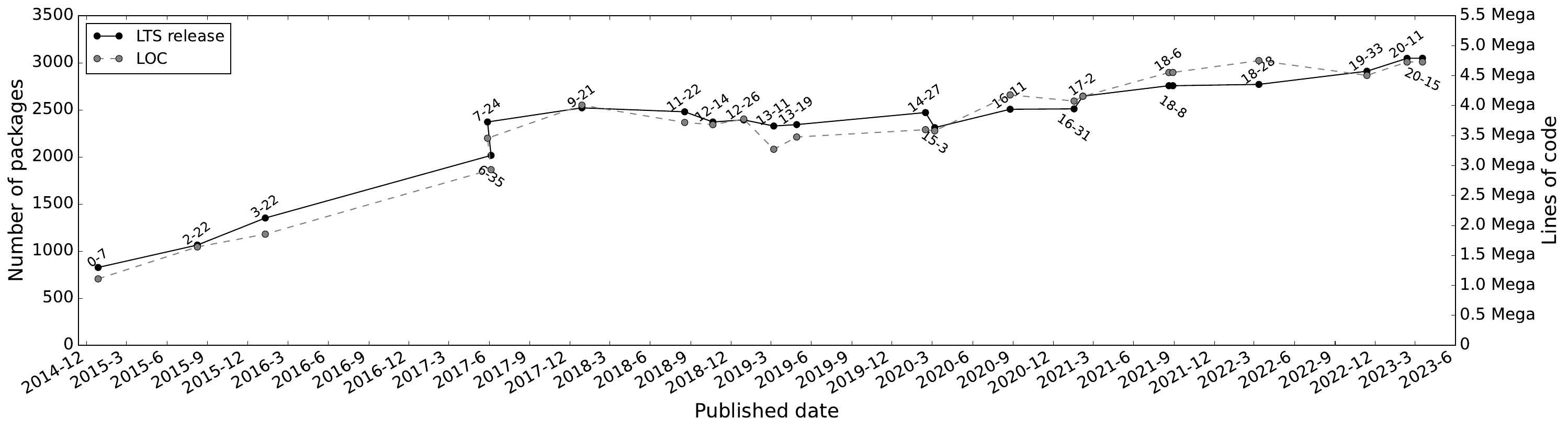}
\caption{\stackage releases}
\label{fig:evolution}
\end{figure*}

\fref{fig:evolution} shows the release evolution during the analyzed time frame, specifically showing 
the number of 
packages and Lines-Of-Code (LOC) per release, and the availability date for the releases. From the 
figure, we can observe that the number of packages and LOC increase proportionally, hence, we next 
refer to both as packages. In addition, we can see how \stackage had a gradual 
growth of packages throughout its first releases, highlighting that there is an interval of one or more 
years among releases until release 6-35. From release 7-24 onwards, the number of packages 
did not undergo significant changes when compared to the first releases. Finally, there are updates 
that are 
made in short intervals of time. For example, we can see that since release 11-22, there are 5 updates 
in approximately one year. 

\begin{figure*}[htbp]
\centering
\begin{subfigure}[t]{0.8\textwidth}
  \centering
  \includegraphics[width=\textwidth]{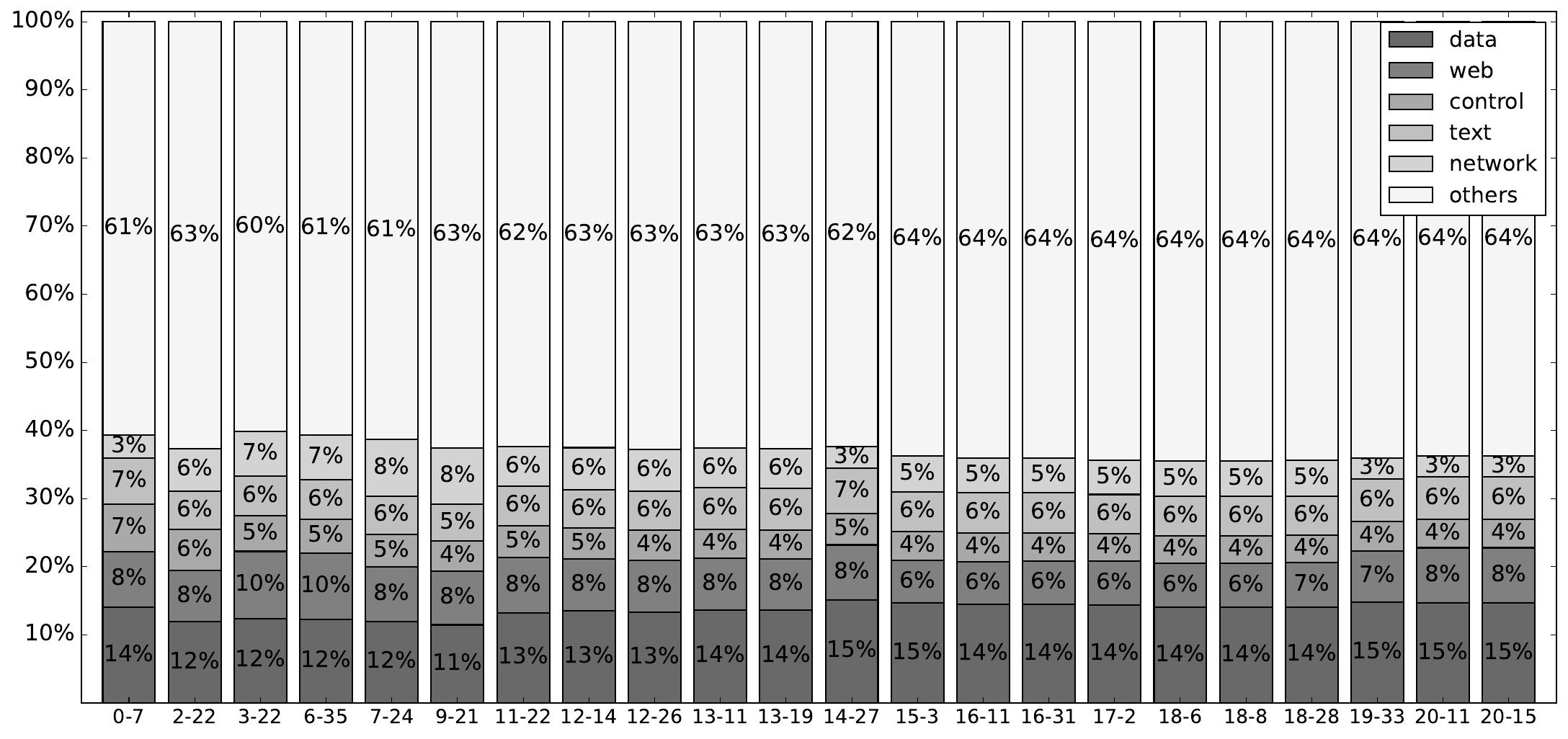}
  \caption{Package percentage by category} 
  \label{fig:dist_by_cat}
\end{subfigure}
\hfill
\begin{subfigure}[t]{0.8\textwidth}
  	\centering
  	\includegraphics[width=\textwidth]{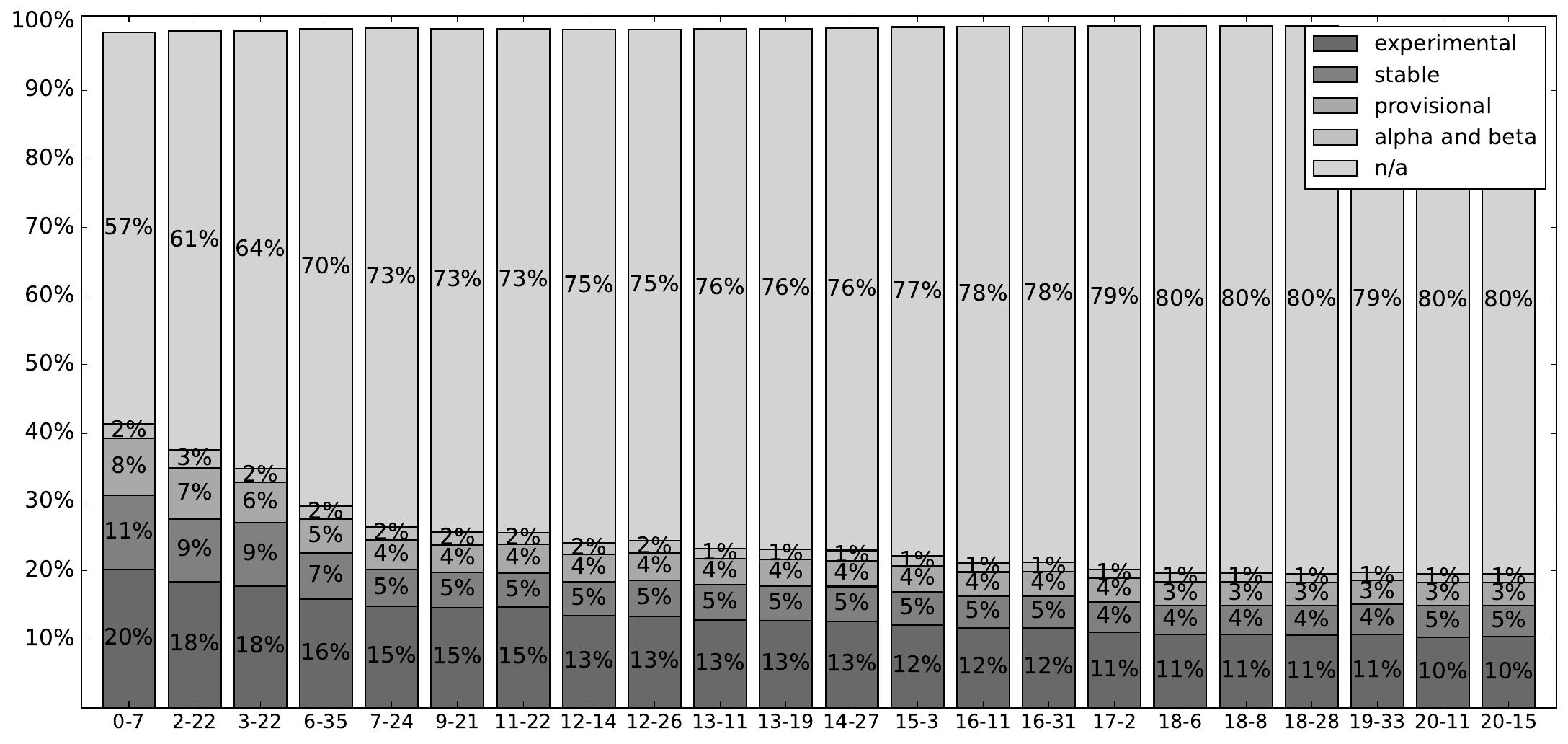}
    \caption{Package percentage by stability level}
  	\label{fig:dist_by_stab}
\end{subfigure}
\caption{Package percentage distribution with respect to category and stability}
\label{fig:pkg_distribution_cat_stab}
\end{figure*}

\fref{fig:pkg_distribution_cat_stab} describes the release evolution through categories and stability of 
packages. \fref{fig:dist_by_cat} presents how the top five categories (\ie \co{data}, \co{network}, 
\co{control}, \co{text}, \co{web}) evolve over different releases. From the figure, we can see how 
some categories decreased their percentage of packages from the first release; for example, the 
\co{web} and \co{control} categories had a decrease of 2\% and 3\% respectively. Other categories, like 
\co{data}, despite varying over time, have a constant average percentage of their presence in the 
\stackage. Finally, the \co{network} category continuously increases its presence in packages over 
time. \fref{fig:dist_by_stab} presents the presence of stable packages over time. From the figure, we 
can observe that throughout its releases, the information in the stability field has decreased notably. 
The consequence of this is that most packages in \stackage do not have a defined type of stability. 
Therefore, most of packages with information in the stability field are classified as \co{experimental}.    


As mentioned in \fref{sec:mppd}, the {\em category} and {\em stability} fields are strings without a 
format. As a consequence, we can find many different categories and stabilities that are not significant 
to show in \fref{fig:pkg_distribution_cat_stab} (\eg empty string). These findings are similar to other 
studies~\cite{figueroa2017,figueroaAl:scp2020}, which analyze the current \hackage repository. The 
reason for the poor results is mainly that developers do not follow a protocol when submitting a new 
package release.   

\subsection{Statistic Analysis}
\label{sec:statisticAna}

To identify genuine package  variance over time, we apply the 
{\em One-Way Anova} test~\cite{ross2017one}, given that our dataset does not follow a {\em normal 
distribution}. Thank to this test, we know whether the hypothesis that the set of packages does not have 
significant variance over the evolution of \stackage is valid. Additionally, we applied a 
\emph{Tukey}~\cite{abdi2010newman} test to specifically evaluate which pair of datasets presents 
significant variance between them over time.

We conducted the One-Way Anova test on all categories, resulting in a $p-value < 0.05$, which implies 
that our dataset exhibits significant differences, thus rejecting the {\em null hypothesis}. Subsequently, 
we apply the Tukey test to the categories. On the one hand, the results showed that the category with 
the most significant variance compared to the others is \co{data}. On the other hand, \co{control} and 
\co{web} did not overlap in their confidence intervals at any point, indicating a significant difference 
between them over time.


We also apply the One-Way Anova and Tukey tests for package stability in \stackage. The Anova 
results with a $p-value < 0.05$ and the Tukey test show that a large portion of stability types exhibit a
significant variance among them, especially the \co{Experimental} stability. In addition, \co{alpha} and 
\co{beta}, being such small datasets, did not show significant differences. This leads us to believe that 
the \haskell community may not effectively distinguish between what constitutes \co{alpha} or \co{beta} 
stability.

%% file: answers.tex

\section{Results: Answering Research Questions}
\label{sec:results}

We now turn our attention to the six research questions posit in this study, 
with the goal of understanding the evolution and use of \stackage and its monad packages. 
To answer these questions we process the information gathered from packages as Pandas 
\co{DataFrames} (\fref{sec:mppd}), as described in \fref{sec:marq}.

\subsection{\fref{enum:imported}.  Which packages are imported the most by other \stackage packages? Do these packages have {\em unstable} or {\em incompatible} dependencies according to their \stackage release?}
\label{sec:arq1}

\begin{figure*}[th]
  \centering
  \includegraphics[width=1\linewidth]{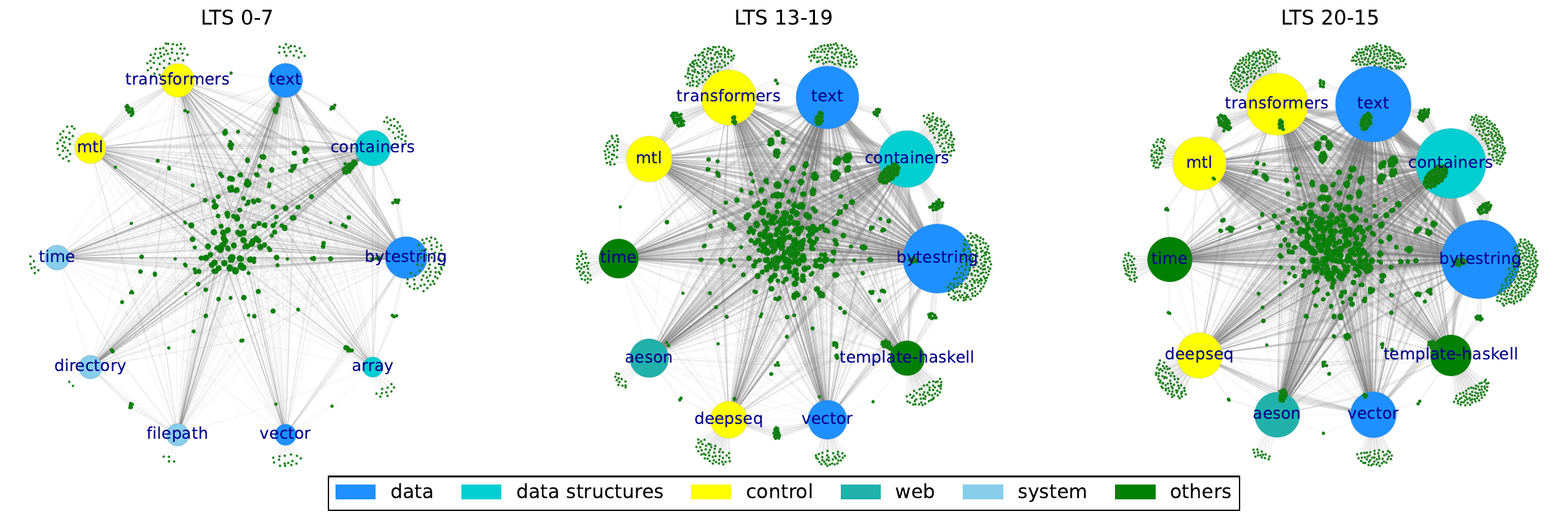}
  \caption{\fref{enum:imported} Packages Dependencies}
  \label{fig:dependencyEvolution}
\end{figure*}

Using the {\em dependencies} field in the built data frames, we created a list of dependencies (with 
their versions) for each package in a release. \fref{fig:dependencyEvolution} shows a dependency 
network graph for three releases of \stackage from the analyzed set: first (2014) to the left, middle (2019), 
and latest (2023) to the right. At a first glance, we can clearly observe that the number of 
dependencies increases with the 
number of new packages in the repository. From the evolution of \stackage, we can observe that 
the top 7 packages keep their dependency ranks, these are: \co{time}, \co{mtl}, \co{transformer}, 
\co{text}, \co{containers}, \co{bytestring}, and \co{vector}. Although the \mtl and \transformers 
packages are not the most used (in terms of dependencies), these packages are widely used in 
the evolution of the repository. Finally, \fref{fig:dependencyEvolution} shows that the categories of the top packages vary. 

\begin{figure*}[t] 
 \begin{subfigure}[t]{0.5\textwidth}
  \centering
  \includegraphics[width=\textwidth]{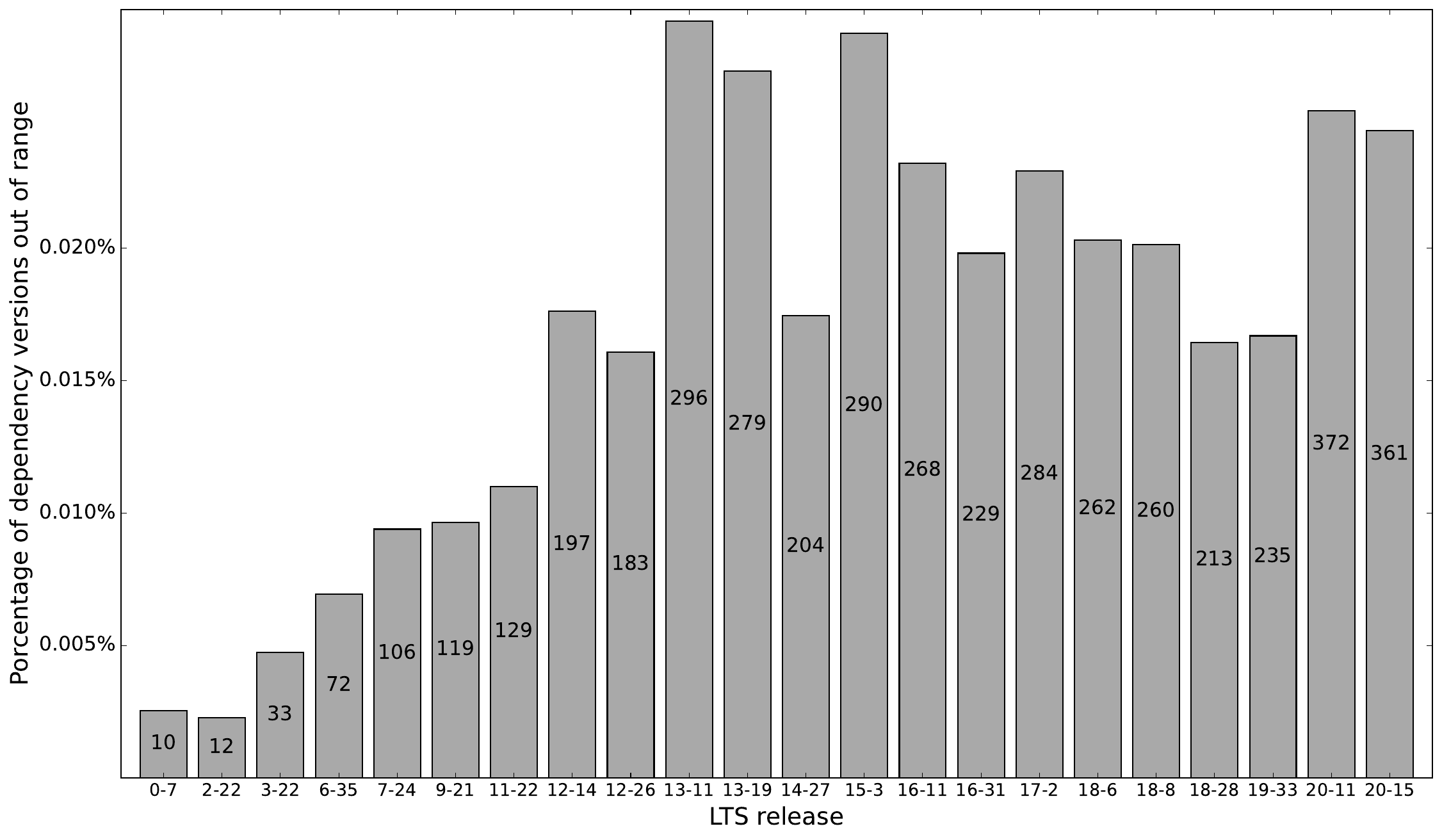}
  \caption{Unstable dependencies per release}
  \label{fig:rq1_2}
\end{subfigure}
\hfill
\begin{subfigure}[t]{0.5\textwidth}
  \centering
  \includegraphics[width=\textwidth]{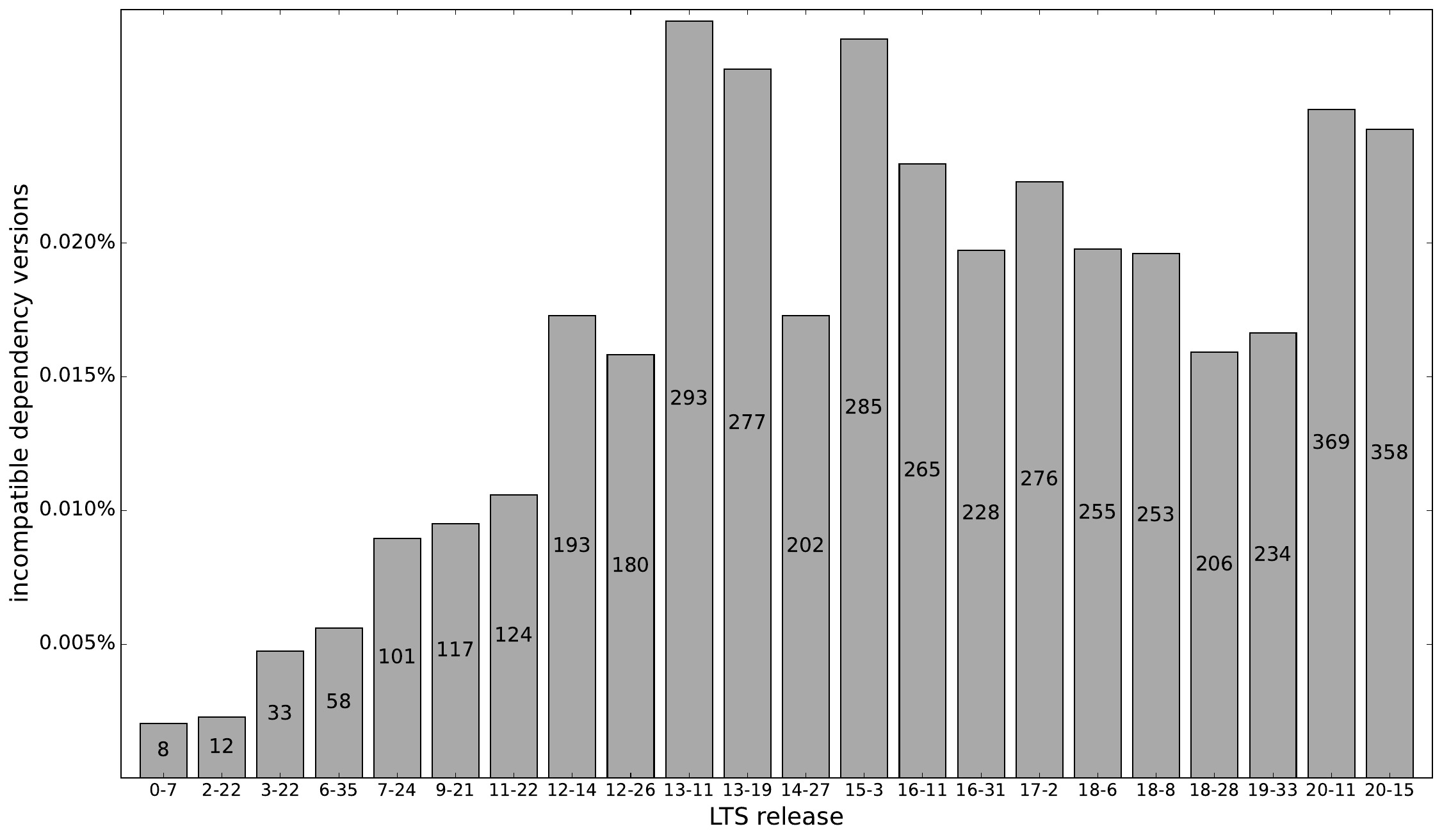}
  \caption{Incompatible dependencies per release}
  \label{fig:rq1_3}
\end{subfigure}
\vspace{4pt} 
\caption{\fref{enum:imported}. Unstable and Incompatible dependencies per \stackage release}
\label{fig:rq1_2_3}
\end{figure*}

When reasoning about the versions of dependencies per package, we present two findings. 
First, \fref{fig:rq1_2} 
shows the number and percentage of packages with dependency versions that are not available in  
\stackage, that is, showing the number and percentage of unstable dependencies for that release. 
While there are not many of these packages, there is a growing trend that might affect the stability 
promise of \stackage in the future. Second, \fref{fig:rq1_3} shows that some packages depend on 
one same package but with 
different version ranges; resulting incompatible dependencies between packages. For this analysis,
we used the first revision of each \co{.Cabal} file. In the following, we 
revisited this analysis using the last revision of \co{.Cabal} files.     

\subsubsection{Revisiting: Unstable and Incompatible Dependencies}
\label{sec:arq12}

\begin{figure*}[ht] 
 \begin{subfigure}[t]{0.5\textwidth}
  \centering
  \includegraphics[width=\textwidth]{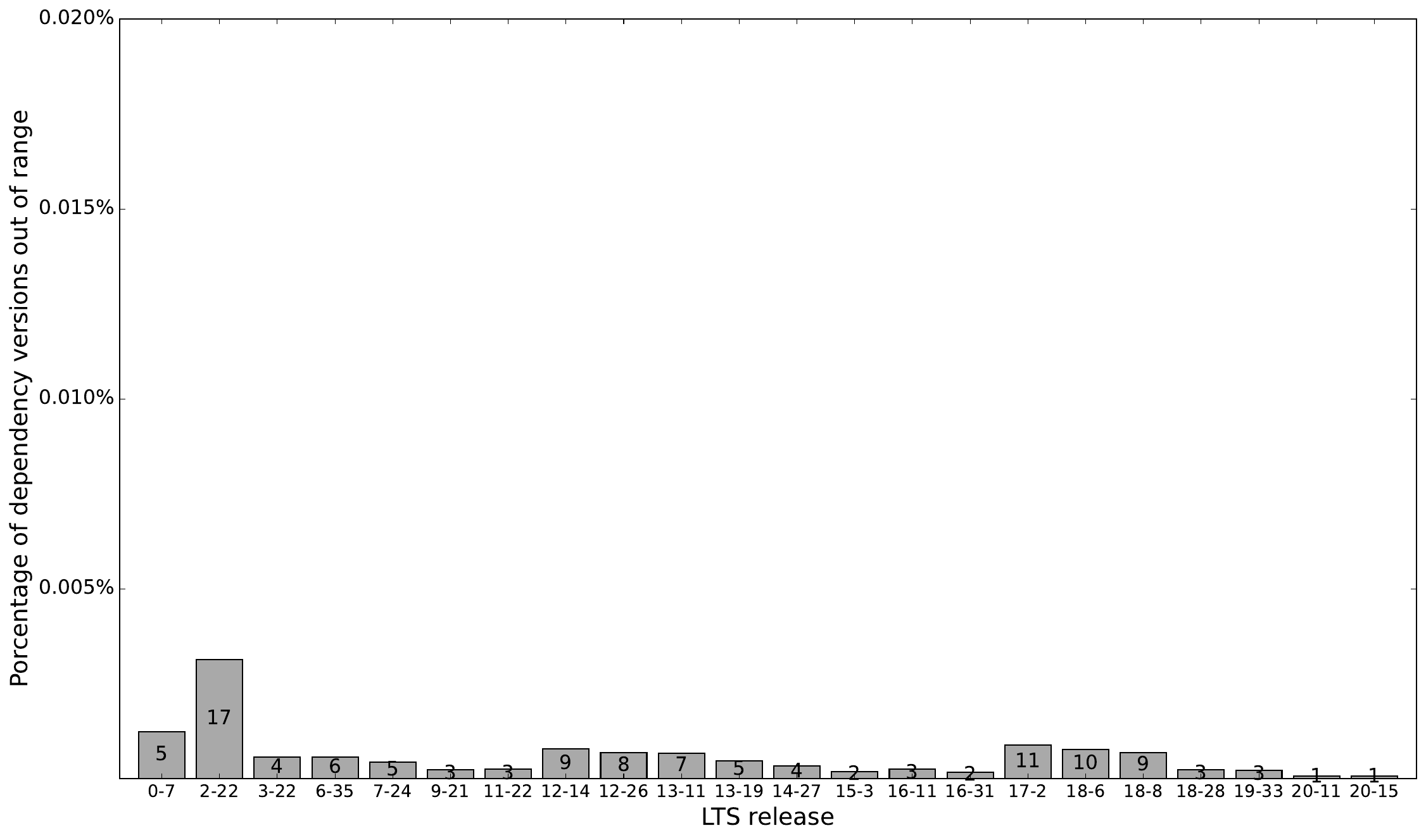}
  \caption{Unstable dependencies per release}
  \label{fig:rq1_2_revised}
\end{subfigure}
\hfill
\begin{subfigure}[t]{0.5\textwidth}
  \centering
  \includegraphics[width=\textwidth]{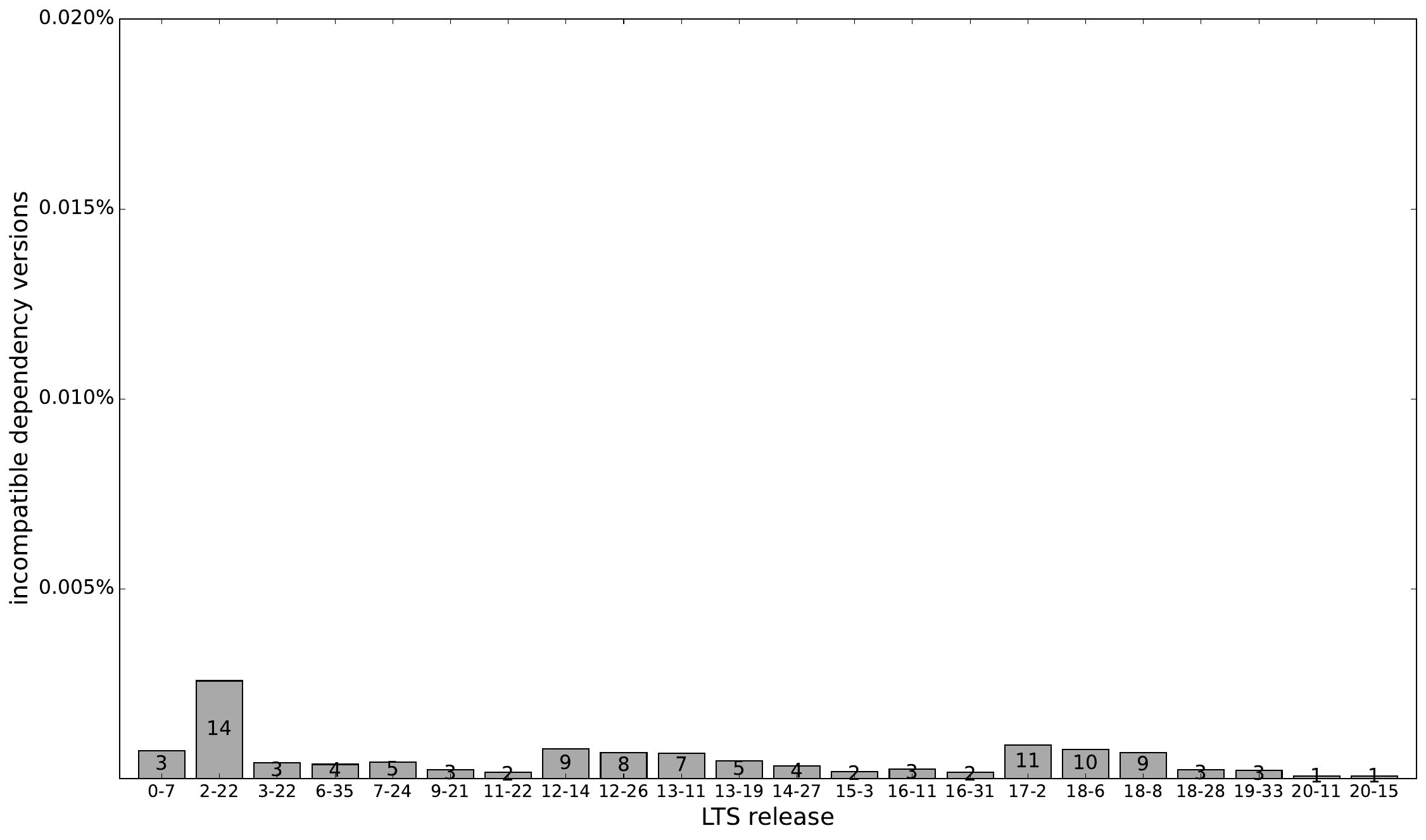}
  \caption{Incompatible dependencies per release}
  \label{fig:rq1_3_revised}
\end{subfigure}
\vspace{4pt} 
\caption{\fref{enum:imported}. Unstable and Incompatible dependencies per \stackage release Using the revised version of \co{.Cabal} files. }
\label{fig:rq1_2_3_revised}
\end{figure*}

In an earlier iteration of our work~\cite{legerAl:sac-se2022}, we shared the results of 
\fref{enum:imported} with the \stackage 
community through a community thread in 2022: ``Safety issues in Stackage?''~\cite{commercialHaskell:2022}. In this thread, we highlighted the issues found in 
\fref{fig:rq1_2_3}. The maintainers and collaborators in \stackage 
remarked that the \co{.Cabal} file of a specific package and release also contains different 
{\em revisions} (beyond versions); which we did not take into account. To clarify this point, we 
extended the analysis to 
include the latest revision of \co{.Cabal} files. However, from the extended analysis,
we can still find a few unstable and incompatible dependencies, as shown in  
\fref{fig:rq1_2_3_revised}. Therefore, we conclude the problem might still persist.

\subsection{\fref{enum:dependencies}. What is the average number of direct and indirect dependencies 
per package?} 
\label{sec:arq2}

\begin{figure*}[hptb]  
\begin{subfigure}[t]{0.5\textwidth}
  \centering
  \includegraphics[width=\textwidth]{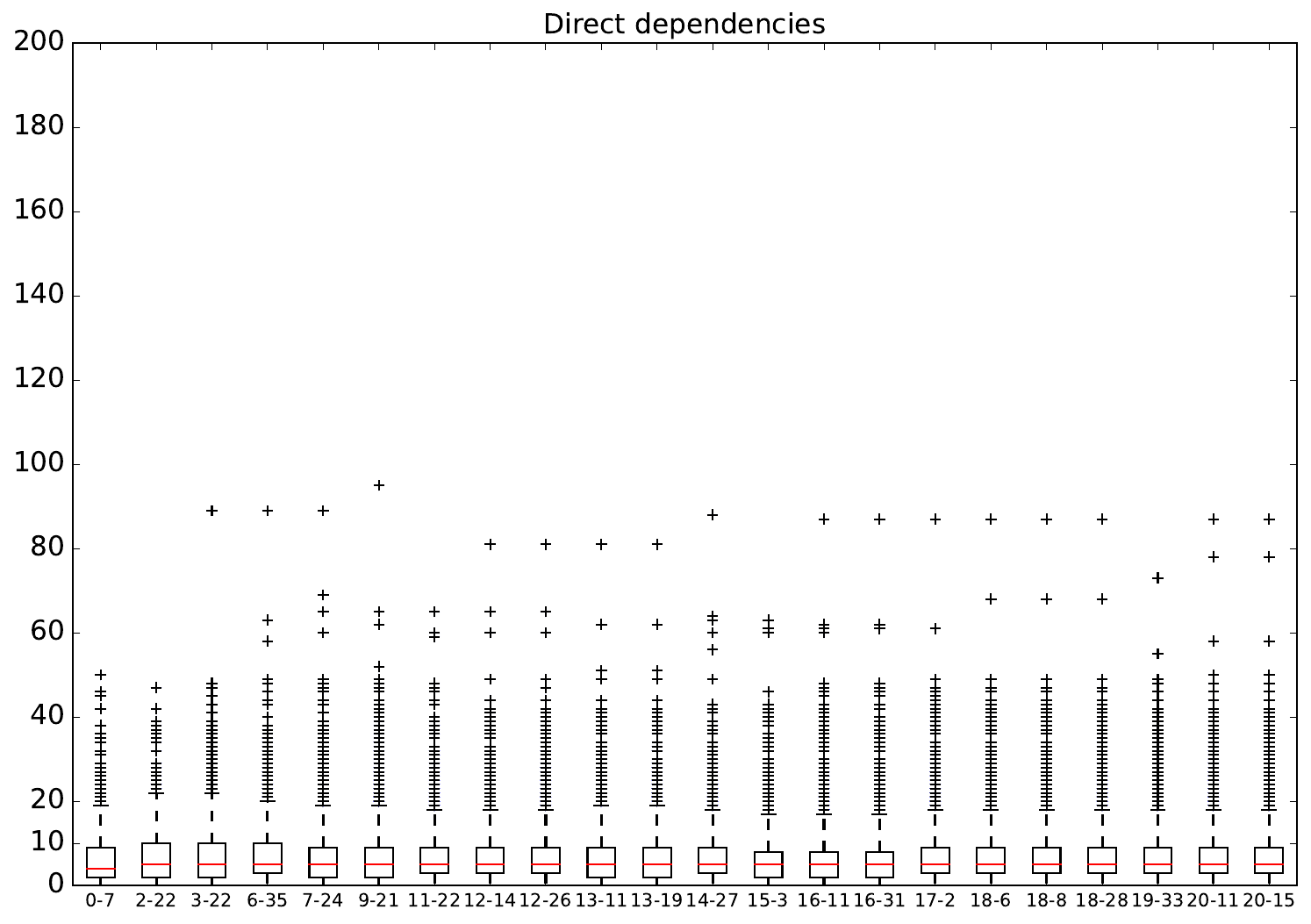}
  \caption{Average direct dependencies}
  \label{fig:rq2_1}
\end{subfigure}
\hfill
\begin{subfigure}[t]{0.5\textwidth}
  \centering
  \includegraphics[width=\textwidth]{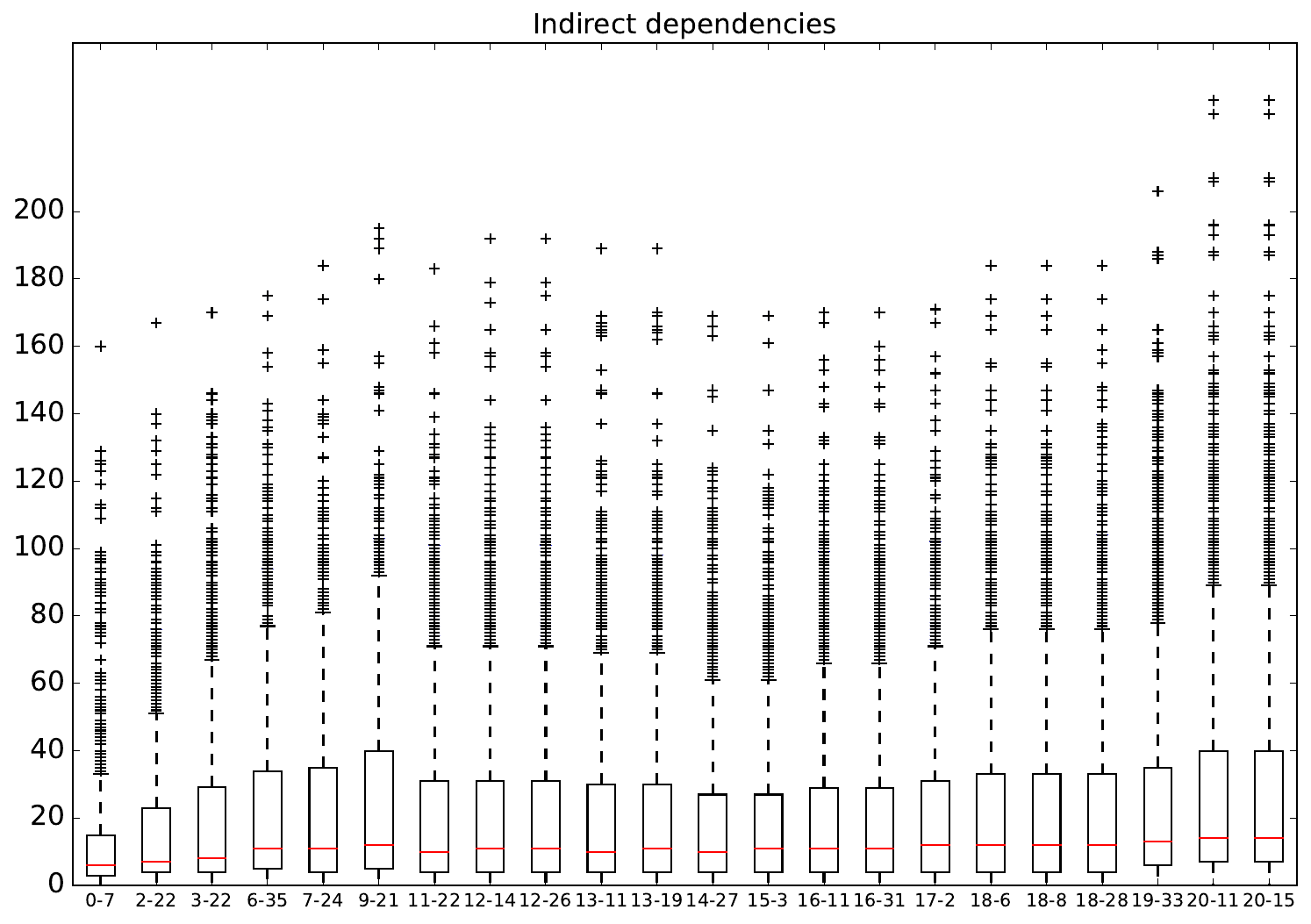}
  \caption{Average indirect dependencies}
  \label{fig:rq2_2}
\end{subfigure}
\vspace{4pt} 
\caption{\fref{enum:dependencies}. Number of package dependencies per release}
\label{fig:indirect-dependencies}
\end{figure*}

To answer this question we use box graphs for direct and indirect dependencies for each release. 
\fref{fig:rq2_1} shows the number of direct dependencies is less than 10, throughout the 
\stackage evolution. Some packages are outside of this 
average, with a higher number of dependencies, going up to 90 for outliers. 
\fref{fig:rq2_2} shows the number of indirect dependencies per package is higher than that of 
direct dependencies. In this case we observe a high variance of the average. Regarding indirect 
dependencies, many packages can be considered outliers, showing that the average (around 10) 
of these dependencies is not representative.   

Overall the average (in)direct dependencies remains constant throughout all releases, hinting that
packages are mostly stable.
The direct dependencies in \fref{fig:rq2_1} are constant across all releases, with an average of 5 to 10 
dependencies per packages. However, we notice a large variance in the number of dependencies, with 
some packages defining between 20 and 50 dependencies. The case of indirect dependencies, in 
\fref{fig:rq2_2}, keeps also constant the average. With a larger sparsity of indirect dependencies, averaging above 10 
dependencies per package, the variance of indirect dependencies is much higher, with some 
packages having between 70 to 140 dependencies.

\subsection{\fref{enum:frequency}. How frequently are packages updated?} 
\label{sec:arq3}

\begin{figure}[htpb] 
  \centering
  \includegraphics[width=0.8\textwidth]{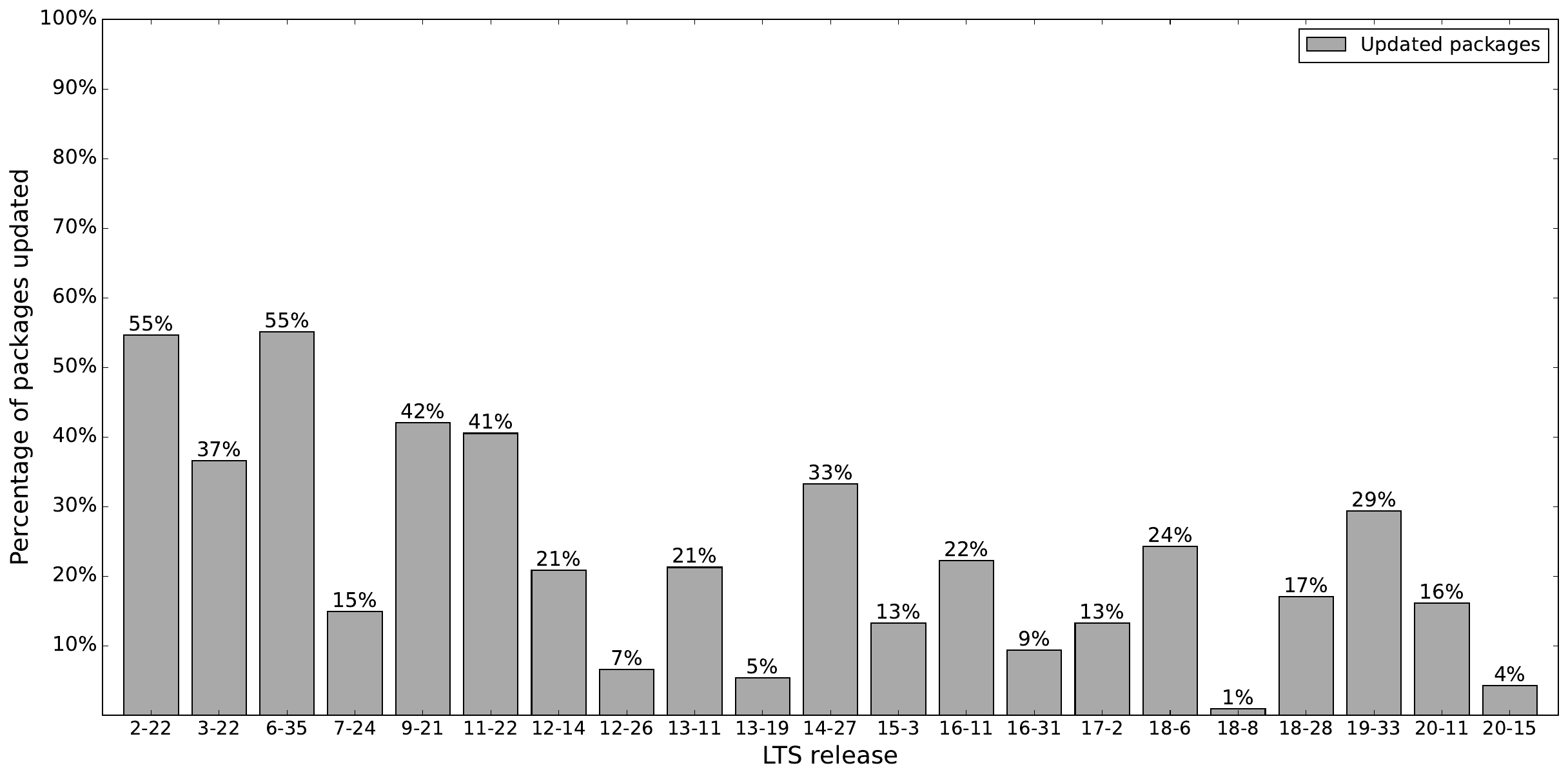}
    \caption{\fref{enum:frequency}. Percentage of updated packages per release}
  \label{fig:rq3}
\end{figure}

To answer this question, we counted the number of packages that updated their versions for each 
release. We then compared the number of {\em updated} packages to the number of all packages in 
a release. \fref{fig:rq3} shows the result of this comparison as the percentage of updated packages 
per release. This result, can be used as a proxy for {\em technical lag}~\cite{gonzalezAl:oss2017} 
in dependencies, which is useful to know how frequently bug fixes and new features are included in 
packages. In the figure, we can see that a higher percentage of packages are updated between 
releases 2-22 and 6-35; there is also a smaller number of updates between releases 12-14 and 
13-19. An explanation for this, is the time interval between releases. In the beginning there is an 
average of 6 months per release, then there is only an average of 3 months per release, therefore 
giving developers less time to update their packages from one \stackage release to the next.

\subsection{\fref{enum:monads-evol}. How have the selected monad packages evolved?} 
\label{sec:arq4}

\begin{figure}[ht]
  \centering
  \includegraphics[width=\textwidth]{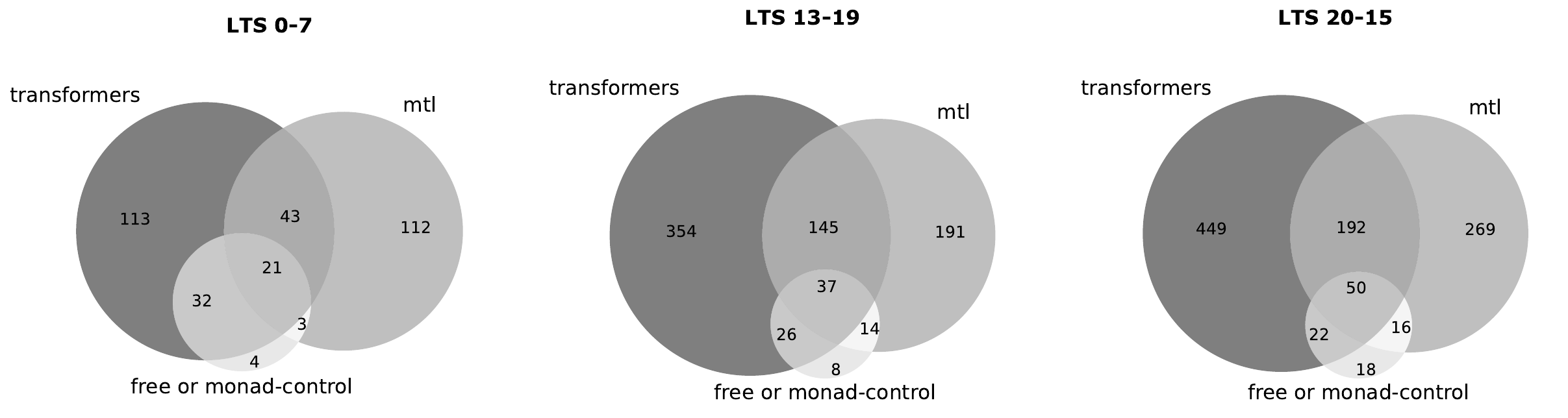}
  \caption{\fref{enum:monads-evol}. Monad package evolution over time}
  \label{fig:rq4}
\end{figure}

To answer this question, we parsed the source code of each package to know which packages import 
at least one of the four monad packages (\mtl, \transformers, \monadcontrol, or \free). \fref{fig:rq4} 
shows the use of each of the monad packages at different moments of the evolution of \stackage 
packages, notably showing the grow in their use, and the combinations across the packages. As we 
can see from the figure, most imports use the \mtl, \transformers monad packages, or a combination 
of the two.

\subsection{\fref{enum:monads-use}. How has the use of the selected monad packages evolved?} 
\label{sec:arq5}

\begin{figure*}[h] 
  \centering
\begin{subfigure}[t]{0.78\textwidth}
  \centering
  \includegraphics[width=\textwidth]{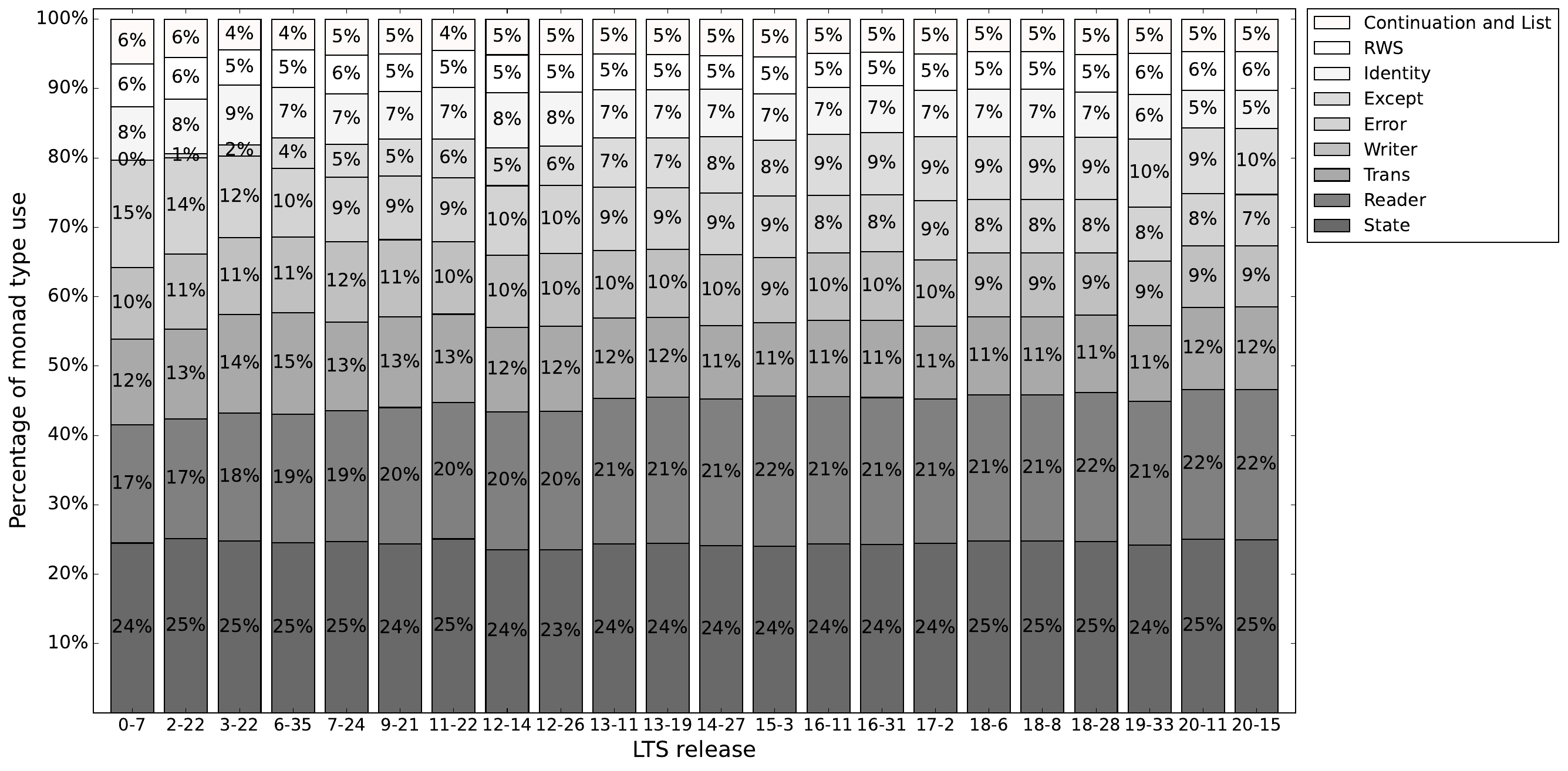}
  \caption{\mtl package evolution usage}
  \label{fig:rq5a}
\end{subfigure}	
~
\begin{subfigure}[t]{0.78\textwidth}
  \centering
  \includegraphics[width=\textwidth]{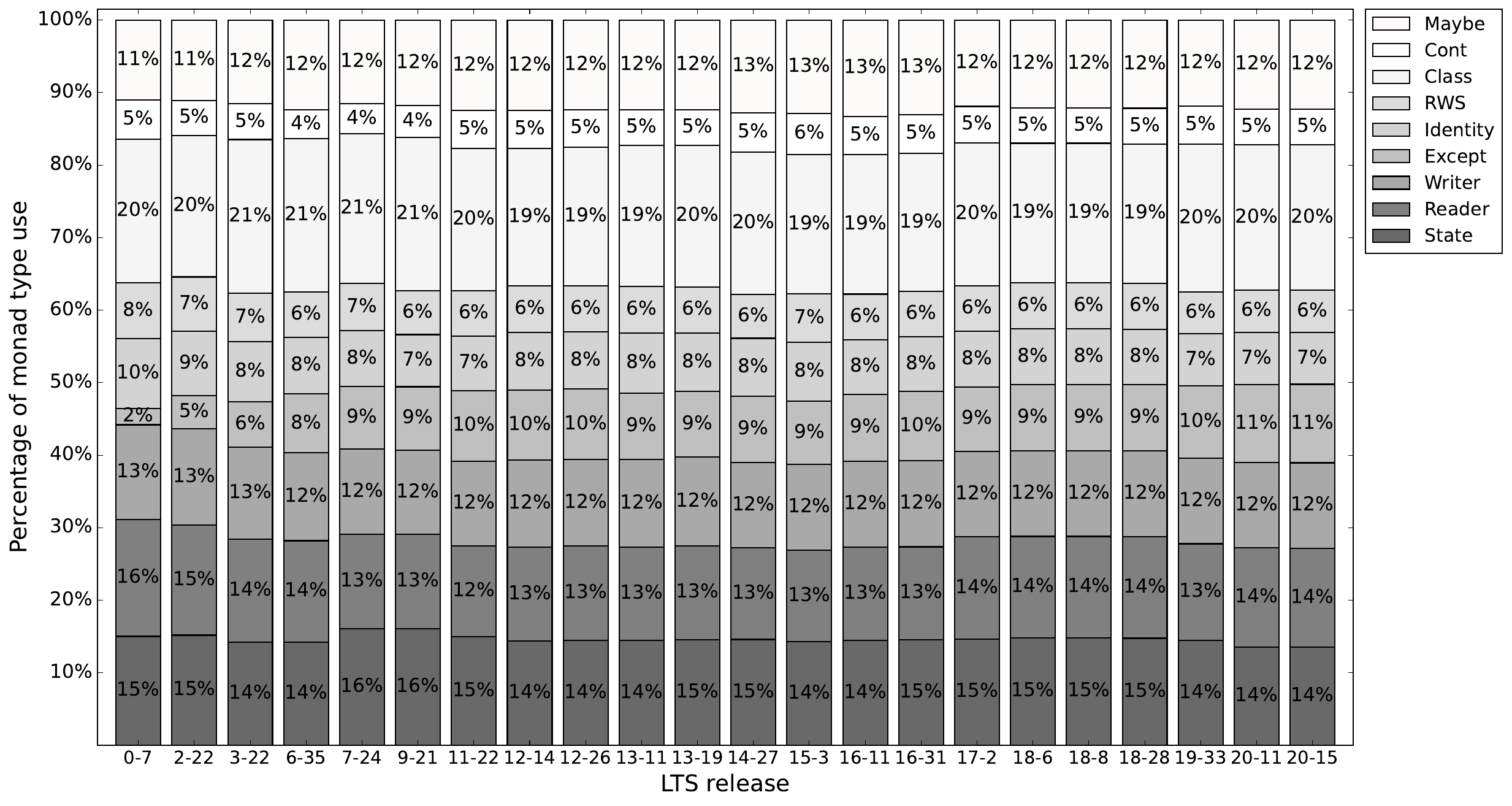}
  \caption{\transformers package evolution usage}
  \label{fig:rq5b}
\end{subfigure}

\begin{subfigure}[t]{0.78\textwidth}
  \centering
  \includegraphics[width=\textwidth]{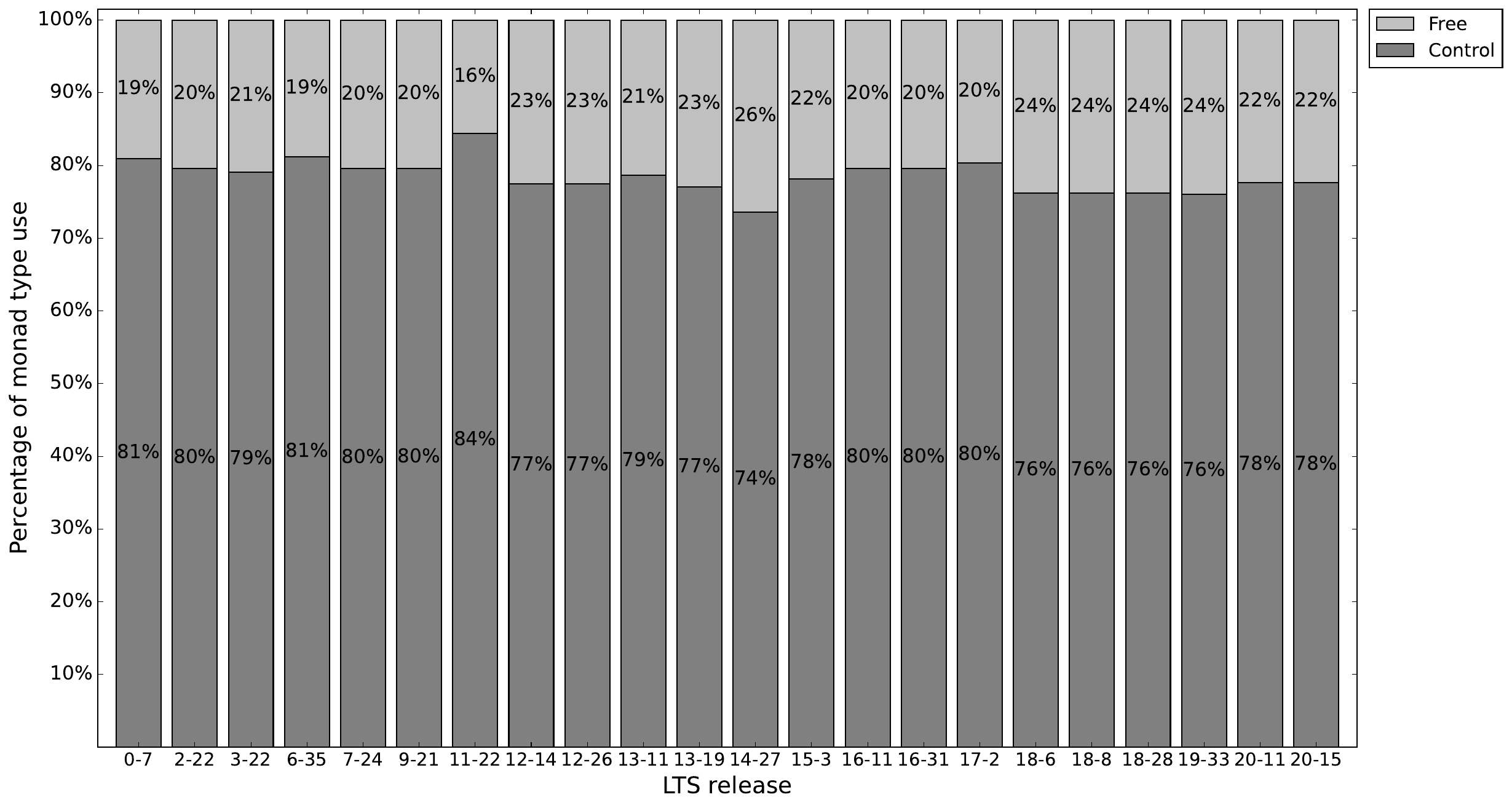}
  \caption{\monadcontrol and \free packages evolution usage}
  \label{fig:rq5c}
\end{subfigure}
\caption{\fref{enum:monads-use}. Usage evolution of monad packages}
\label{fig:rq5}
\vspace{4pt} 
\end{figure*}

To answer this question, we count the amount of imported modules, from the selected monad 
packages (\mtl, \transformers, \monadcontrol, and \free), in each release. \fref{fig:rq5a} and 
\fref{fig:rq5b} show the use of different monad modules in \mtl and \transformers respectively. For 
\mtl and \transformers, we can see that the \co{State} and \co{Reader} monads have a strong 
presence against other monads. This presence may be due to both monad packages allowing 
developers to express stateful behavior. Particularly, in \mtl, \fref{fig:rq5a} shows that the use 
of \co{Error} and \co{Except} are inversely proportional, possibly due to \co{Except} being able to 
work as a replacement for \co{Error} in various scenarios. Regarding \transformers, the use of its 
modules approximately remains stable since release 6-35. Because of the limited participation of 
\monadcontrol and \free, \fref{fig:rq5c} shows the participation distribution of both packages, where 
\monadcontrol is used more than \free.

\subsection{\fref{enum:mtl-transformers-dependencies}. How many packages that depend on the 
\mtl and \transformers packages are added to and removed from \stackage? How many packages 
that depended on these monad packages stopped their dependencies?} 
\label{sec:arq6}

\begin{figure*}[ht] 
 \begin{subfigure}[t]{0.5\textwidth}
  \centering
  \includegraphics[width=\textwidth]{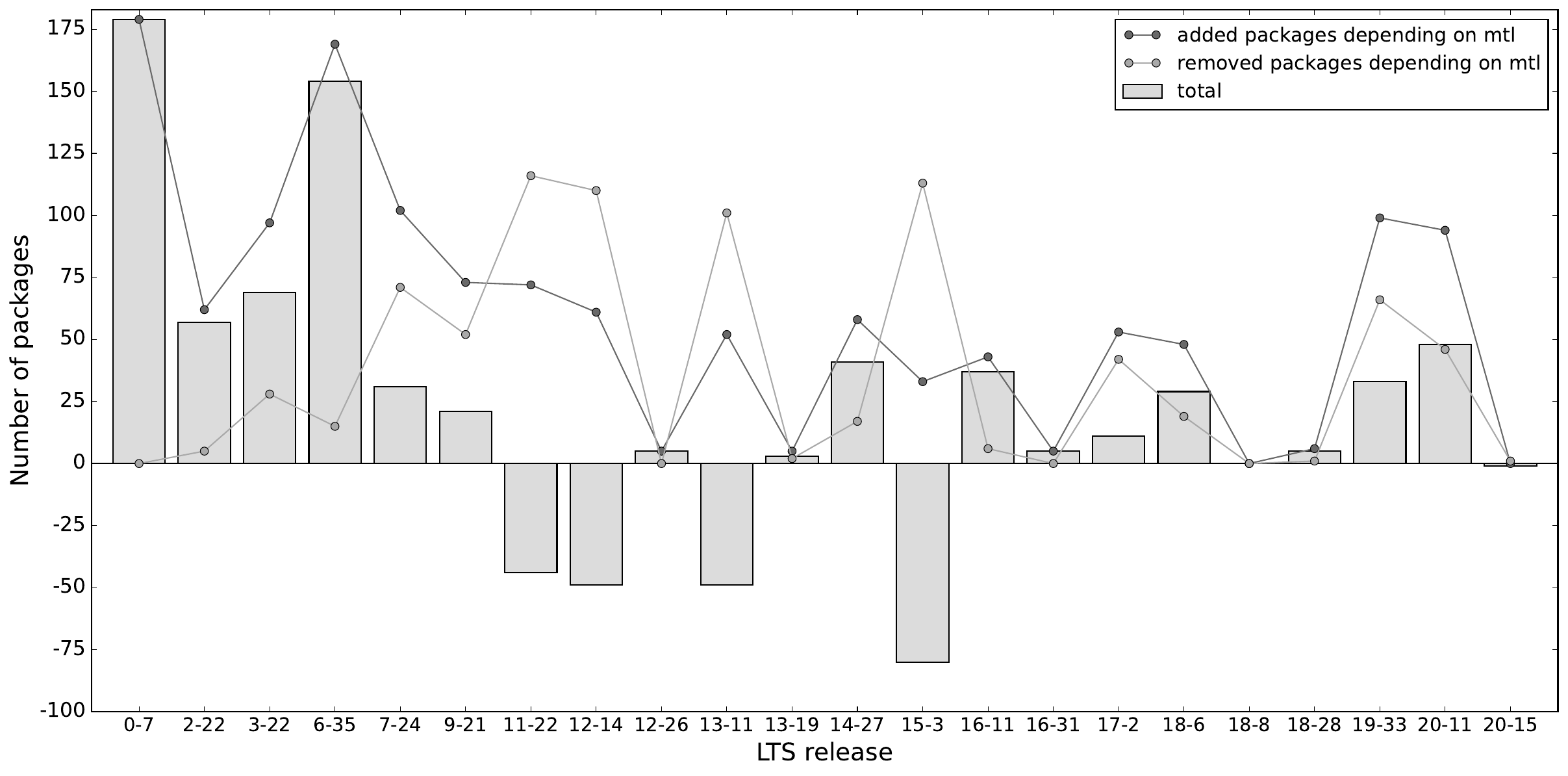}
  \caption{\mtl dependent packages}
  \label{fig:rq6_1_mtl}
\end{subfigure}
\hfill
\begin{subfigure}[t]{0.5\textwidth}
  \centering
  \includegraphics[width=\textwidth]{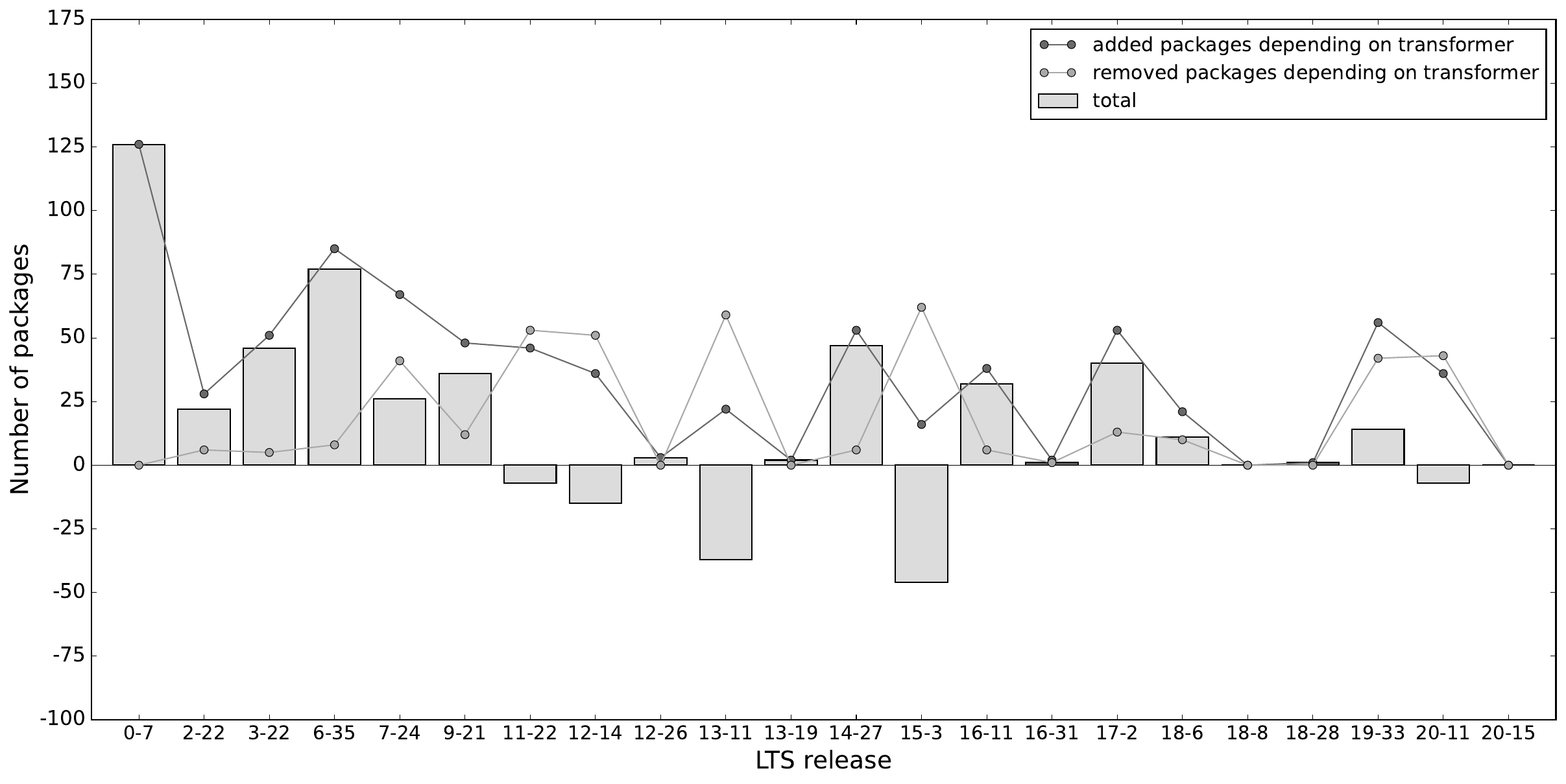}
  \caption{\transformers dependent packages}
  \label{fig:rq6_1_transformers}
\end{subfigure}
\vspace{4pt} 
\caption{Number of added and removed packages dependent on \mtl or \transformers}
\label{fig:rq6_1}
\end{figure*}

\fref{fig:rq6_1} addresses the number of added and removed packages taking into account the 
releases that import \mtl (\fref{fig:rq6_1_mtl}) and \transformers (\fref{fig:rq6_1_transformers}) 
modules. Note that we leave \monadcontrol and \free out of the analysis given their limited 
participation in \stackage. 
For both packages, the initial releases a higher number of packages add \mtl or \transformers than 
the latter ones. Between releases 7-24 and 12-14, we can see a considerable fluctuation of 
packages that are added and removed.

\begin{figure*}[] 
 \begin{subfigure}[t]{0.5\textwidth}
  \centering
  \includegraphics[width=\textwidth]{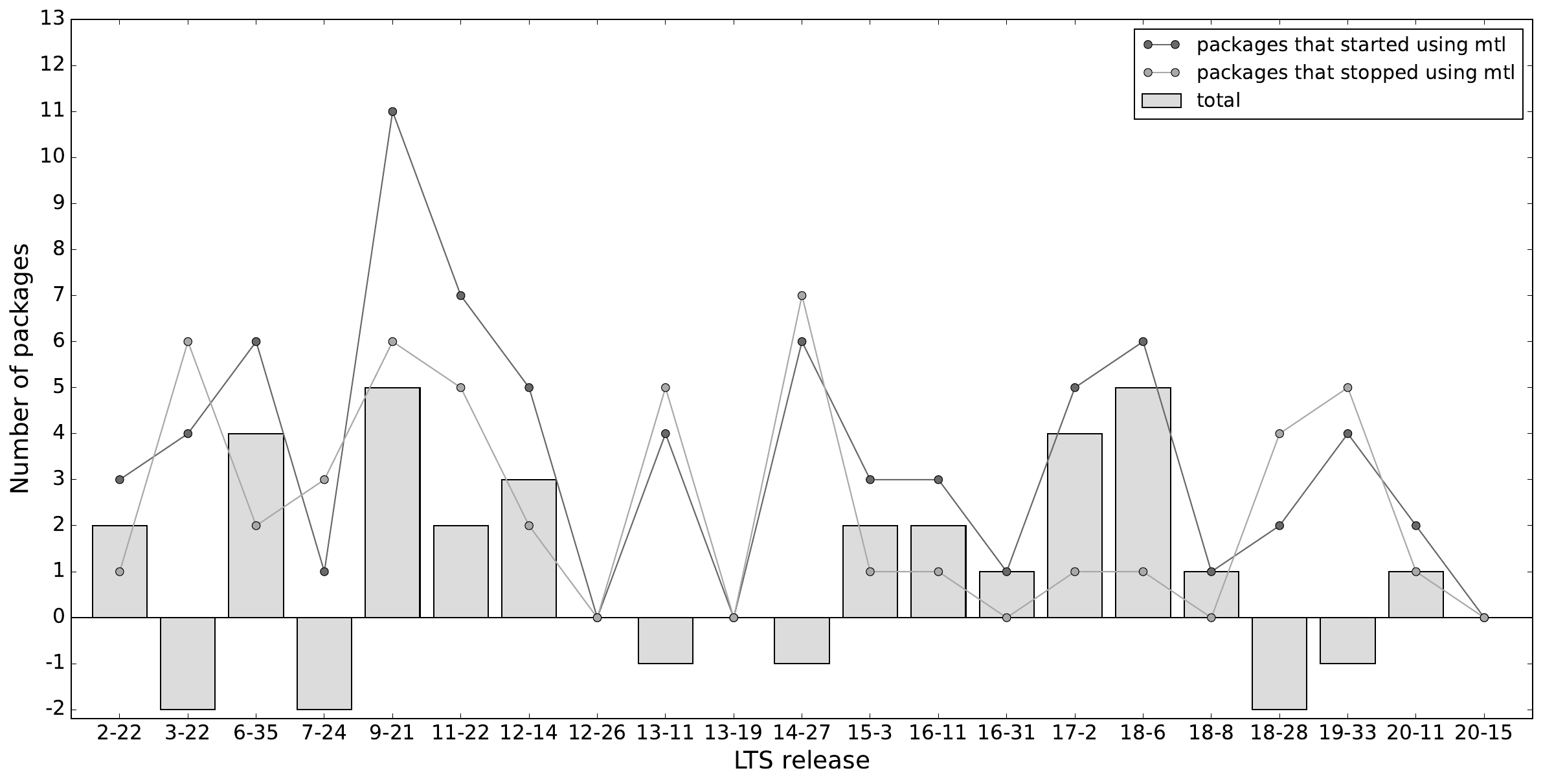}
  \caption{Stopped dependencies to \mtl}
  \label{fig:rq6_2_mtl}
\end{subfigure}
\hfill
\begin{subfigure}[t]{0.5\textwidth}
  \centering
  \includegraphics[width=\textwidth]{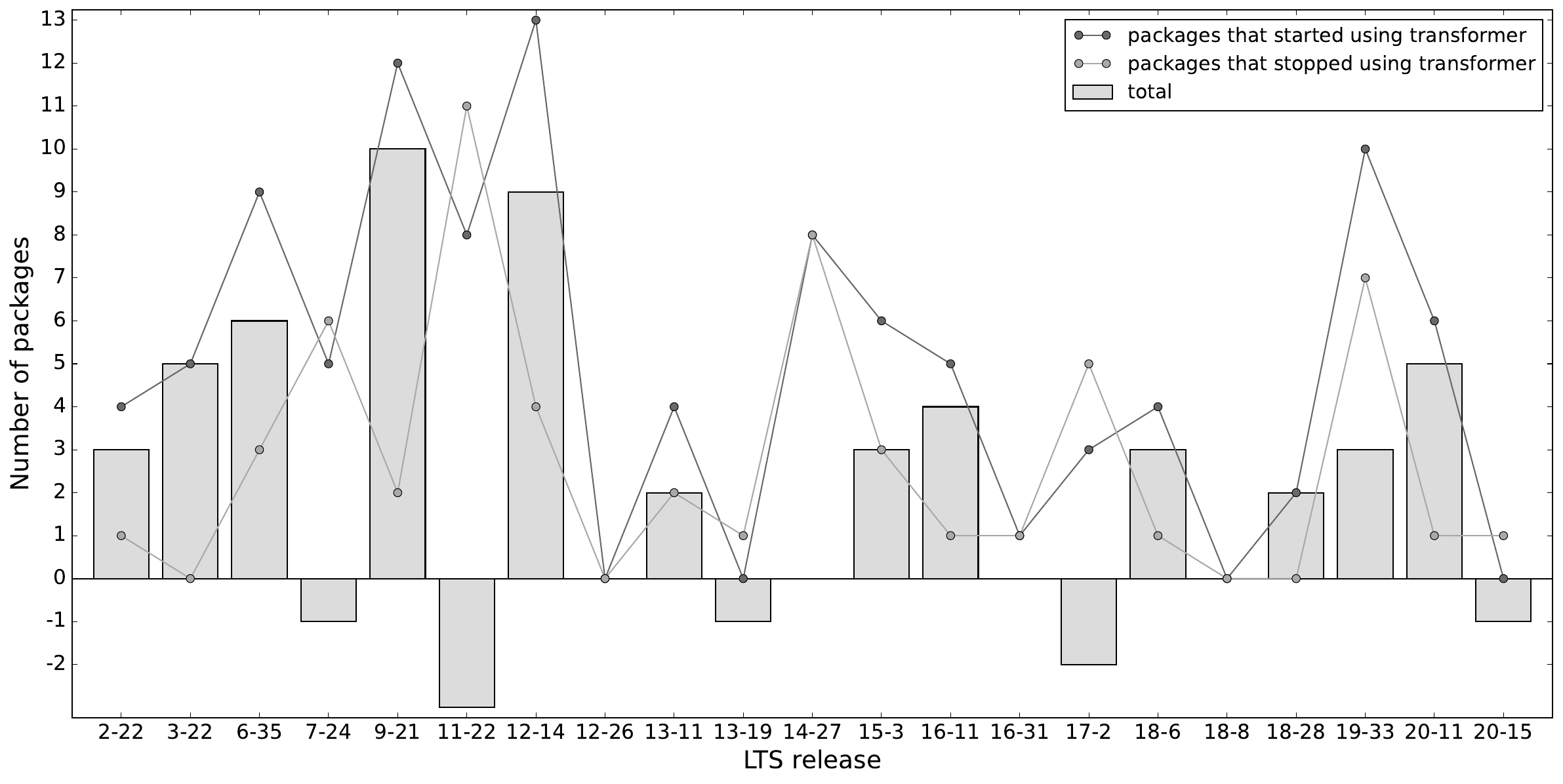}
  \caption{Stopped dependencies to \transformers}
  \label{fig:rq6_2_transformers}
\end{subfigure}
\vspace{4pt} 
\caption{Number of packages that stopped dependencies to \mtl or \transformers}
\label{fig:rq6_2}
\end{figure*}

\fref{fig:rq6_2} shows the number of packages that start and stop using \mtl (\fref{fig:rq6_2_mtl}) and 
\transformers (\fref{fig:rq6_2_transformers}) packages. In the figures, it is easy to realize that a 
maximum of 10 packages ($\approx 0.5\%$ of a release) that did not use \mtl or \transformers, 
started using them later (and vice versa).    

Both the \mtl and \transformers packages present similar behavior as their complementary usage 
brings different package benefits. For example, {\em package compatibility} with other packages 
in \stackage that already use both packages (\eg support tools like QuickCheck).

\subsection{Summary}
\label{sec:summary}

Across the answers to the six research questions, we can highlight the following insights about 
\stackage and its evolution. First, two monad packages (\transformers and \mtl) are among the top 10 
packages most imported. Second, \stackage packages contain incompatible and unstable 
dependencies, even if we consider the {\em revisions} of packages. Third, the average (in)direct 
dependencies has been similar throughout all releases. Fourth, there is an apparent trend in the lack 
of package updates. Fifth, apart from \transformers and \mtl, no other monad packages are widely 
used. Sixth, when a package starts using a monad package (\transformers or \mtl), it is not common 
to use other monad package later (and vice versa). Finally, developers do not follow a standard for the 
{\em stability} field in a package; indeed, over 50\% is empty.


%% file: related.tex

\section{Related Work}
\label{sec:rw}

We follow guidelines from the \ac{MSR}~\cite{hemmatiAll:msr2013} community to carry out the 
exploratory study of \stackage. As this article has shown, that through empirical studies carried out 
on the repository packages, it is possible to know the identity that it has been forging throughout its 
evolution. This section discusses studies about the evolution of repositories for different programming 
languages first, and then studies focusing on \haskell.

\citet{Chinthanet2021} conduct a study on the unstable dependencies in \npm~\cite{npm}, a repository 
for \javascript. The goal study is to analyze the lag in packages that have vulnerable dependencies. By 
lag in dependencies, the authors refer to the period of time since an error is detected in the dependency 
until this error is solved. ~\citet{kikasAl:msr2017} explain how updating or eliminating dependencies 
affects a repository that strongly depends on specific packages, giving rise to potential errors in the 
repository. The study is carried out in the repositories of \javascript, \ruby, and \rust by making a 
network of their dependencies to analyze their evolution. Finally, ~\citet{Plakidas2017}, present an 
exploratory and descriptive study of packages in the \cran repository of the R programming language 
to give a vision of the distribution and the identity that R has for the developer community.

Regarding \haskell, we can find a few studies that explore the language community in a known 
repository. \citet{figueroa2017,figueroaAl:scp2020} explore how the developer community uses different 
kinds of monads in \hackage.  ~\citet{Bezirgiannis2013} use \hackage to research how developers use 
generic programming language constructs, tools, and libraries in \haskell. 
Finally,~\citet{morris:haskell2010} uses \hackage to discover programming pattern in practice. We note
that unlike our work, none of these studies analyze the evolution of the repositories over time.

%% file: conclusion.tex

\section{Conclusion and Future Work}
\label{sec:conc}

Programming language communities are usually build around repositories that allow developers to 
access and share packages. Normally, the relevance of repositories in the community is associated 
to their {\em stability}. As an example,  \stackage is a curated repository for stable \haskell packages in 
the \hackage repository. In \stackage, maintainers make available LTS releases that contain a set of 
packages with specific versions. Although \stackage is widely used (in industry, as its main target), we 
are not aware of many empirical studies on how this repository has evolved, including the use of its 
defining characteristic in \haskell: monads. As a consequence, there is a lack of potential guidelines 
that \stackage maintainers may follow. To the best of our knowledge, this paper is the first large-scale 
analysis of the evolution of the \stackage repository: 22 Long-Term Support~(LTS) releases 
(in the period 2014 - 2023). The main findings of our work highlight that: first, there is a growing number 
of packages that 
\begin{enumerate*}[label=(\arabic*)]
\item depend on packages whose versions are not available in a release, and 
\item have dependencies with incompatible versions; however, the revisions of \co{.Cabal} files of 
packages aim to resolve most of these issues. 
\end{enumerate*}
Second, the \mtl and \transformers packages are the top 10 packages most used in \stackage. 
Third, like previous studies~\cite{figueroa2017,figueroaAl:scp2020}, the repository maintainers should 
be careful when providing unstructured fields for metadata such as {\em stability} or \emph{category} 
because developers do not follow a protocol to provide this kind of information. Finally, for \mtl and 
\transformers, the use of the \co{State} and \co{Reader} monad seems to be the most used in \stackage. 

The findings presented in this paper are firmly based on the analysis of dependencies and modules 
imported in package releases. This kind of analysis obfuscates a fine-grained use of functions 
provided by modules in packages, including monads. This is because this analysis does not allow us 
to know what and how imported functions are really being used (if they are). We plan to carry out a 
complete static analysis of source code to answer previous questions. As a result, this analysis may 
provide us with insights about why specific packages (\eg \mtl) are being imported.